\documentclass{pasj00}
\usepackage{ulem}
\usepackage{color}
\draft

\begin{document}
\SetRunningHead{Tsai et al.}
	{Molecular Superbubbles and Outflows from NGC 2146}
\Received{2008 December 15}
\Accepted{2009 April 25}

\title{
Molecular Superbubbles and Outflows from the Starburst Galaxy NGC 2146
}

\author{An-Li \textsc{Tsai},\altaffilmark{1,2}
	Satoki \textsc{Matsushita},\altaffilmark{2}
	Kouichiro \textsc{Nakanishi},\altaffilmark{3}
	Kotaro \textsc{Kohno},\altaffilmark{4}\\
	Ryohei \textsc{Kawabe},\altaffilmark{3}
	Tatsuya \textsc{Inui},\altaffilmark{5}
	Hironori \textsc{Matsumoto},\altaffilmark{5}\\
	Takeshi G. \textsc{Tsuru},\altaffilmark{5}
	Alison B. \textsc{Peck},\altaffilmark{6}
	\and
	Andrea \textsc{Tarchi},\altaffilmark{7,8}
	}
\altaffiltext{1}{Department of Earth Sciences,
	National Taiwan Normal University, \\
	No.88, Dingzhou Rd. Sec. 4, Taipei 11677, Taiwan}
	\email{altsai@asiaa.sinica.edu.tw}
\altaffiltext{2}{Academia Sinica, Institute of Astronomy and Astrophysics,
	P.O.~Box 23-141, Taipei 10617, Taiwan}
\altaffiltext{3}{Nobeyama Radio Observatory, Minamimaki, Minamisaku,
	Nagano 384-1305}
\altaffiltext{4}{Institute of Astronomy, University of Tokyo,
	2-21-1 Osawa, Mitaka, Tokyo 181-0015}
\altaffiltext{5}{Department of Physics, Faculty of Science,
	Kyoto University, Sakyo-ku, Kyoto 606-8502}
\altaffiltext{6}{Joint ALMA Office, Av El Golf 40, Piso 18, Santiago, Chile}
\altaffiltext{7}{INAF - Osservatorio Astronomico di Cagliari,
	Loc. Poggio dei Pini, Strada 54, 09012 Capoterra, Italy}
\altaffiltext{8}{INAF - Istituto di Radioastronomia, via Gobetti 101,
	40129 Bologna, Italy}


\maketitle

\begin{abstract}
We present results from a deep (1$\sigma$ = 5.7~mJy~beam$^{-1}$ per
20.8~km~s$^{-1}$ velocity channel) \atom{C}{}{12}O(1-0) interferometric
observation of the central $60\arcsec$ region  of the nearby edge-on
starburst galaxy NGC 2146 observed with the Nobeyama Millimeter Array
(NMA).
Two diffuse expanding molecular superbubbles and one molecular outflow are
successfully detected.
One molecular superbubble, with a size of $\sim1$~kpc and an expansion
velocity of $\sim50$~km~s$^{-1}$, is located below the galactic disk; a
second molecular superbubble, this time with a size of $\sim700$~pc and an
expansion velocity of $\sim35$~km~s$^{-1}$, is also seen in the
position-velocity diagram; the molecular outflow is located above
the galactic disk with an extent $\sim2$~kpc, expanding with a
velocity of up to $\sim200$~km~s$^{-1}$.
The molecular outflow has an arc-like structure, and is located at the
front edge of the soft X-ray outflow.
In addition, the 
kinetic 
energy ($\sim 3 \times 10^{55}$ erg) and the pressure
($\sim 1 \times 10^{-12\pm1}$~dyne~cm$^{-2}$) 
of the molecular outflow is
comparable to or smaller than that of the hot thermal plasma, suggesting that
the hot plasma pushes the molecular gas out from the galactic disk.
Inside the $\sim1$ kpc size molecular superbubble, diffuse soft X-ray
emission seems to exist.
But since the superbubble lies behind the inclined galactic disk,
it is largely absorbed by the molecular gas.
\end{abstract}

\section{Introduction}
\label{sect-intro}

A starburst galaxy is characterized by strong star forming activity.
Starburst phenomena usually occur in the central regions of galaxies, and
produce large numbers of massive stars on short time scales
with star formation rates tens or hundreds of times higher than those in
normal galaxies.
Through stellar winds and energetic supernova explosions, these massive
stars generate a large amount of mechanical energy that can sweep the
surrounding interstellar medium (ISM) from starburst regions, create ISM
bubbles, and generate high velocity galactic winds.
This hypothetical starburst evolution is supported by several numerical
models (e.g., \cite{tom88,str00,mar05}), and is actually based on
various observational results.
Shell-like structures and outflows have been observed in many starburst
galaxies in optical emission lines and soft X-ray emission
(e.g., \cite{leh96,mar98,dah98,str04}).
Shell-like structures are also often observed in nearby galaxies in the
atomic hydrogen (H\emissiontype{I}) line (e.g., \cite{ten88}).

Evidence for these shell-like or outflowing
features from starburst galaxies are, however, rarely seen using molecular
tracers.
The famous cases are the molecular outflows in M82 \citep{nak87} and
NGC 891 \citep{han92}, and molecular bubbles in NGC 3628 \citep{irw96}.
The reasons for the rarity of such detections are mostly their
diffuse and extended nature and the lack of sensitive detections 
in this wavelength range.
Recent improvements of various instruments at millimeter-wave 
have resulted in progress in this field:
The interferometric \atom{C}{}{12}O(1-0) observations toward M82 succeeded in
imaging molecular superbubbles and outflows
(e.g., \cite{nei98,wei99,mat00,wei01}),
and revealed a self-induced starburst mechanism \citep{mat05}.
Molecular superbubbles and/or outflows have also been detected in NGC 253
\citep{sak06}, NGC 4631 \citep{ran00}, and NGC 4666 \citep{wal04}.
Since stars form from molecular gas, observing these peculiar features in
molecular gas can provide information about 
the regulation and feedback processes of
the starburst phenomena.
Here we report a new example of molecular superbubbles and outflows from
the starburst galaxy NGC 2146.

NGC 2146 is a nearby (17.2 Mpc, 1$\arcsec$ = 80 pc; \cite{tul88}),
IR luminous ($1.2 \times 10^{11}$~L$_\odot$; \cite{san03}),
edge-on ($i = 63\arcdeg$; \cite{del99}) starburst galaxy.
Its optical appearance shows disturbed features with peculiar spiral
arms, which cannot be explained by a simple model of spiral structure
\citep{ben75,gre06}.
The neutral hydrogen images display 100~kpc scale tidal tails, although
there is no evidence of a nearby companion galaxy \citep{fis76,tar01}.
A large amount of molecular gas is concentrated around the central region,
which is enough to support the starburst activities \citep{jac88,you88}.
The central region hosts several compact radio continuum sources,
identified as supernova remnants, radio supernovae, and ultra-compact
and/or ultra-dense H\emissiontype{II} regions \citep{tar00}.
The soft X-ray images show kpc-scale outflows from the starburst region
around the galactic center \citep{inu05,del99,arm95}.
All these characteristics are reminiscent of those present in the nearby
starburst galaxy M82, where kpc-scale outflows have also been detected,
together with a molecular superbubble.
Given the similarity between the two galaxies, it is natural to expect that
similar features can be revealed also in NGC 2146.
Therefore, we have performed deep observations in the \atom{C}{}{12}O(1-0) line
toward the central region of NGC 2146 as a case study for molecular
superbubbles and outflows.

\section{Observations and Data Reductions}
\label{sect-obs}

We performed deep \atom{C}{}{12}O(1-0) observations toward the central
$60\arcsec$ region of the edge-on starburst galaxy NGC 2146 with the
Nobeyama Millimeter Array (NMA), which consists of six 10-meter antennas
located at the Nobeyama Radio Observatory (NRO).
The observations were made during 2001 November to 2002 April with three
configurations.
The total on-source time is $\sim$44 hours.
The phase tracking center of our observation is
$\alpha$(J2000) = \timeform{6h18m37s.6},
$\delta$(J2000) = \timeform{78D21'24".1}.
We used tunerless SIS receivers \citep{sun94}, and observed
the \atom{C}{}{12}O(1-0) line in the upper side band.
Double side band system temperature was about 200 -- 300~K for most of the
observations.
The backend used was the XF-type spectro-correlator Ultra Wide Band
Correlator (UWBC; \cite{oku00}), configured to have 512~MHz bandwidth
with 256 channels (i.e., a 2~MHz channel width or 5.2~km~s$^{-1}$ velocity
resolution).
We observed 3C279 as a bandpass calibrator, and 0633+734 as an amplitude
and phase calibrator.
The flux scale of 0633+734 was determined by comparisons with Mars,
Uranus, and Neptune, and the uncertainty in the absolute flux scale is
estimated to be $\sim10\%$.

The data were calibrated using the NRO software package ``UVPROC II''
\citep{tsu97}, and were CLEANed using standard procedures 
implemented in the NRAO software {\it AIPS}.
The maps were made with natural weighting, 
with a final synthesized beam
size of $3\farcs4 \times 2\farcs8$ (280~pc $\times$ 230~pc) 
and a position angle of $108\arcdeg$.
The noise level of the full spectral resolution maps (5.2~km~s$^{-1}$) 
is 11.4~mJy~beam$^{-1}$.
The noise level of the 20.8~km~s$^{-1}$ 
resolution channel maps in figure~\ref{ch0112} is 5.7~mJy~beam$^{-1}$.

\section{Results}
\label{sect-res}

\subsection{Overall Molecular Gas Distribution and Kinematics}
\label{sect-overall}

In figure~\ref{m0m1}, we present the \atom{C}{}{12}O(1-0) integrated intensity
(moment 0, figure~\ref{m0m1}a) and the intensity-weighted mean velocity
field (moment 1, figure~\ref{m0m1}b) maps of the central region of
NGC 2146.
Most of the CO emission is concentrated along the galactic disk with the
inner CO intensity peaks of the image roughly following the major axis,
but with the outer regions of the galactic disk clearly tilted
counter-clockwise from the major axis.
These features are consistent with previous interferometric \atom{C}{}{12}O(1-0)
observations \citep{jac88,gre00,gre06}.
The velocity field (figure~\ref{m0m1}b) shows that the overall
velocity contours are roughly symmetric, although here, as well,
a tilt is present at the outer part of the galactic disk.
The position angle of this galaxy, derived from its kinematics (i.e.,
from the velocity field), is 137$\arcdeg$ (an angle of 0$\arcdeg$
corresponds to north, increasing counterclockwise).
This value is consistent with that derived from past CO observations
(135$\arcdeg$; \cite{dum01}).
Figure~\ref{ch0112} shows the \atom{C}{}{12}O(1-0) channel maps of the same
region with a velocity resolution of 20.8~km~s$^{-1}$.
The overall molecular gas features along the galactic disk
(P.A.\ of $137\arcdeg$)
are similar to those often seen in galaxies with flat-rotation curves.
In addition, there are some diffuse structures outside the galactic disk.

The position-velocity ($p-v$) diagram taken along the major axis
(figure~\ref{pv_aspx}) can be fitted in the central region of the galactic
disk by two rigid body rotation curves.
The rigid body rotation curve between $-4\arcsec$ and $+4\arcsec$ shows
a steeper gradient
than that between $-10\arcsec$ and $+10\arcsec$.
Since the former structure is located closer to the nucleus of the galaxy,
hereafter we refer to it as the ``nuclear disk'', 
while the latter is referred to as the ``inner disk''.
The region outside the inner disk (hereafter called the ``outer disk'')
shows flat rotation.

The total CO flux density detected in figure~\ref{m0m1}a is measured as
$\sim1.8\times10^{3}$~Jy~km~s$^{-1}$ (= 110~K~km~s$^{-1}$).
The CO luminosity over a $1\times10^{7}$~pc$^{2}$ area is then calculated as
$1.1\times10^{9}$~K~km~s$^{-1}$~pc$^{2}$.
The CO luminosities observed with the IRAM 30~m single dish telescope and
the Plateau de Bure Interferometer are
$(1.2\pm0.1)\times10^{9}$~K~km~s$^{-1}$~pc$^{2}$ over
$1\times10^{7}$~pc$^{2}$ area and $0.8\times10^{9}$~K~km~s$^{-1}$~pc$^{2}$
over $0.8\times10^{7}$~pc$^{2}$ area, respectively (with their values
re-scaled because of the different distance for NGC 2146 adopted by us;
\cite{gre06}).
If we allow for a calibration error of 10\% in our data, our CO luminosity
is comparable with the single dish luminosity reported in \citet{gre06},
indicating that our observations recovered all the CO flux, and hence, are
not affected by the missing flux problem.

\subsection{Large-Scale Expanding Molecular Superbubble and Outflow}
\label{sect-large}

\begin{figure*}
\begin{center}
\FigureFile(80mm,80mm){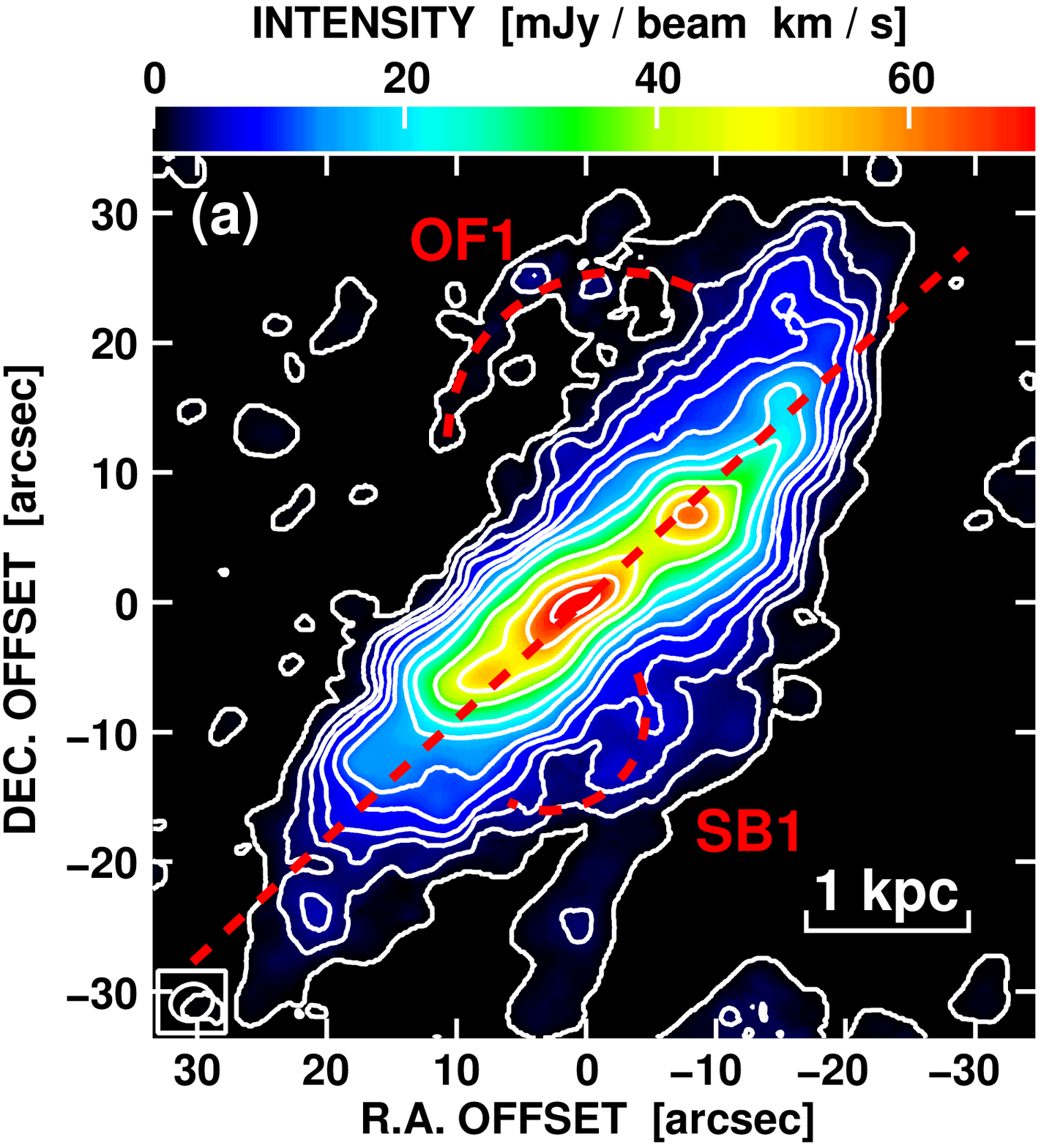}
\FigureFile(80mm,80mm){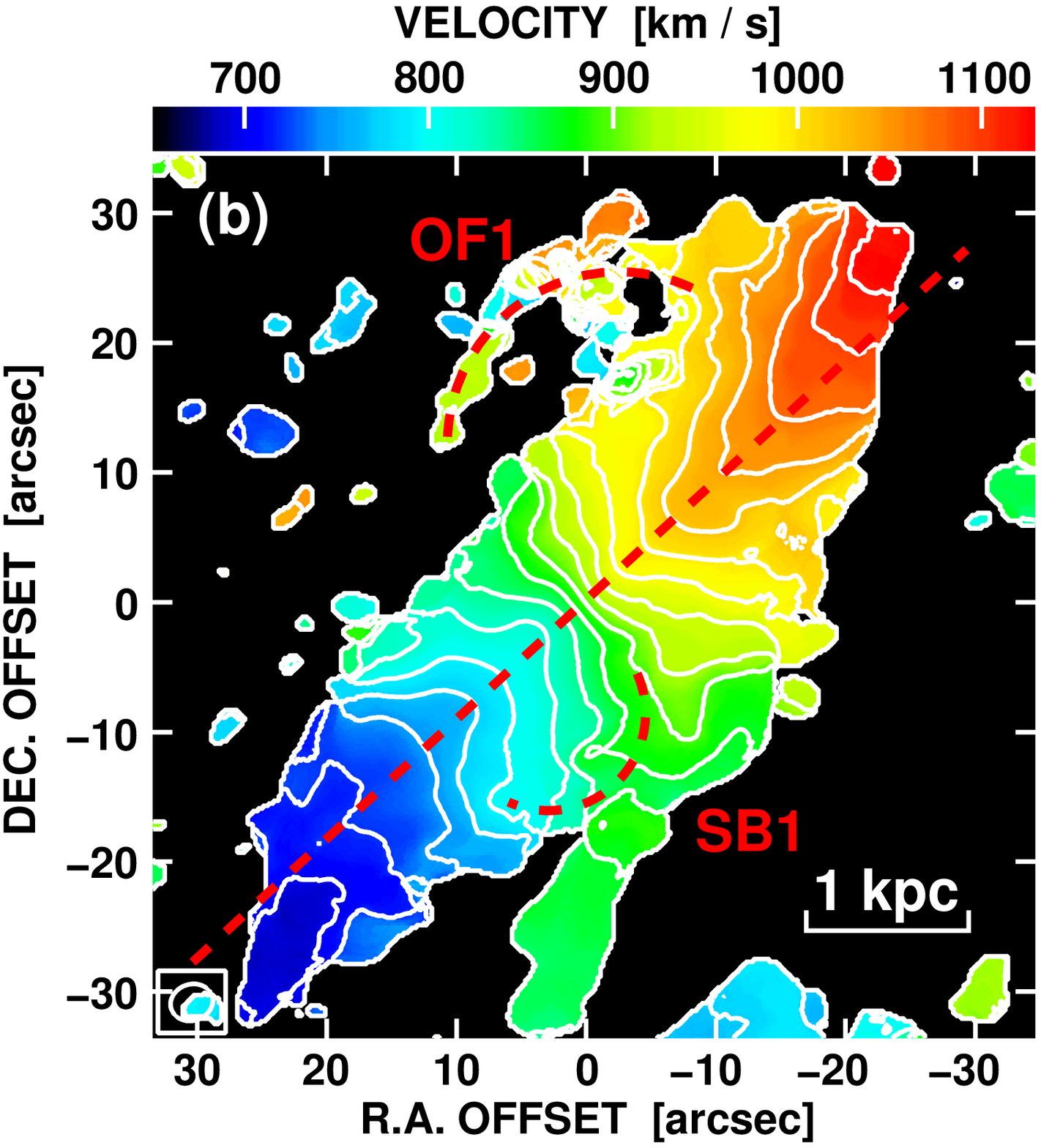}
\end{center}
\caption{
\atom{C}{}{12}O(1-0) integrated intensity (moment 0) and intensity weighted mean
velocity field (moment 1) maps of NGC 2146.
The central position corresponds to R.A.\ = \timeform{6h18m37s.6}
and Dec.\ = \timeform{78D21'24".1} (J2000). 
The synthesized beam size is $\timeform{2".8} \times \timeform{3".4}$,
which is shown at the bottom left corner of each figure.
The red straight dashed lines indicate the position of the major axis
(P.A.\ = 137$\arcdeg$).
The {\it SB1} label indicates the position of an expanding superbubble,
and the arc {\it OF1} shows an outflow,
possibly a remnant of an expanding superbubble.
The linear structure elongated from {\it SB1} toward the south is associated
with a dust lane or ``dust finger'' (see Sect.~\ref{sect-res} for
details).
(a) \atom{C}{}{12}O(1-0) moment 0 map.
    The contour levels are (0.1, 4, 7, 10, 15, 20, 30, 40, 60, 80, 100,
    and 120) $\times$ 585~mJy~beam$^{-1}$~km~s$^{-1}$
    (= 5.69~K~km~s$^{-1}$).
(b) \atom{C}{}{12}O(1-0) moment 1 map.
    The contour levels are from 695, 720, 745, $\dots$, and
    1120~km~s$^{-1}$, increasing with 25~km~s$^{-1}$.
}
\label{m0m1}
\end{figure*}

In addition to the bright features seen in the galactic disk region,
we also detected diffuse and extended
structures above and below the galactic disk.
These features are labelled {\it OF1} and {\it SB1} in figure~\ref{m0m1}a.
In figure~\ref{m0m1}b, these regions show velocity structures more
disturbed than the disk region, and obvious deviation from symmetric
velocity contours can be easily seen.

Figure~\ref{ab} is a $p-v$ diagram
along the minor axis after the molecular gas data have been integrated
along the major axis (see figure~\ref{m0rot43} for the range over which
the integration has been performed).
The $p-v$ diagram shows that most of the molecular gas is concentrated in
the central ``vertical'' region at the position of zero arcsec, which
corresponds to the galactic disk.
Outside of the disk, however, the two diffuse and extended
structures, {\it SB1} and {\it OF1} are also evident.
The arc {\it SB1} shows a half shell-like structure encompassing a cavity,
which corresponds to the {\it SB1} in figure~\ref{m0m1}a.
A similar half shell-like molecular structure seen in the $p-v$ diagrams
of M82 has be been explained in terms of an expanding bubble
\citep{nei98,wei99,wil99}.
Comparable behavior has often been reported in expanding
H\emissiontype{I} superbubbles \citep{deu90}.
The similarities between the distribution and kinematics of {\it SB1}
and those of the expanding bubbles reported in literature lead us to
conclude that {\it SB1} is indeed an expanding molecular superbubble.
From the integrated intensity map (figure~\ref{m0m1}a) and the
major-axis-integrated $p-v$ diagram (figure~\ref{ab}), we can estimate the
radius of {\it SB1} to be $\sim10\arcsec - 15\arcsec$, which corresponds
to $\sim800-1200$~pc at a distance of 17.2 Mpc.
The range of the radius depends on where we assume the center of the
bubble is located; 
the minimum radius corresponds to the edge of the disk, and the
maximum the center of the disk.
The expansion velocity can also be measured from the $p-v$ diagram.
We assume that the superbubble expansion is traced by the intensity peaks
marked by the red dashed line of figure~\ref{ab}.
In this case, the measured expansion velocity is
$\sim50\pm10$~km~s$^{-1}$.

Following the arrow overlapped to the extended region {\it OF1}
(seen as an arc in figure~\ref{m0m1}a), a velocity gradient is seen.
Similar features with linear increasing velocity are often seen in
outflows from young stellar objects in star forming regions in
our Galaxy.
Because of the location, structure, and kinematics of the arc {\it OF1},
we conclude that {\it OF1} is most likely an outflow.
This linear feature in figure~\ref{ab} 
extends from the center of the
galaxy to a position $\sim25\arcsec$ distant from it, hinting at an
extension of the outflow from the galactic center of $\sim2$~kpc.
Near the galactic disk, both blueshifted and redshifted 
components can be seen 
(within $\pm150$~km~s$^{-1}$,
between two red solid lines in the right side of figure~\ref{ab}).
The outflow velocity is increasing linearly from
V$_{LSR}$ of $\sim0$~km~s$^{-1}$ to $\sim+200$~km~s$^{-1}$.

From {\it SB1} in figure~\ref{m0m1}b, a diffuse linear structure with a
very small velocity gradient is pointing toward the south.
This structure can also be clearly seen in the channel maps
(figure~\ref{ch0112}) at a velocity range between 829.8~km~s$^{-1}$ and
850.6~km~s$^{-1}$.
This component matches with the CO clumps reported in \citet{kar04}, and
is probably associated with a dust lane or ``dust finger'' stretching
southwest from the galactic center \citep{you88,hut90,gre00}.

\subsection{Smaller-Scale Molecular Superbubbles}
\label{sect-small}

Together with the two aforementioned diffuse structures, which are obvious
in the integrated intensity map and the $p-v$ diagram, there is another
bubble-like structure detected around the galactic disk region.
The molecular gas in normal circular galactic rotation would be 
concentrated along a rigid- or flat-rotation curve.
Gas not exhibiting this behavior can be considered as gas rotating under
non-circular motion or moving under different mechanisms.

Figure~\ref{pv12chmaj} shows the $p-v$ diagrams along the major axis with
different offsets.
Most of the molecular gas is indeed concentrated along the rigid- or
flat-rotation curves, but some diffuse structures can be seen away from
these curves.
All of the diffuse structures appear only to the west of the
galactic center with its offset between $-5\farcs0$ and $-16\farcs0$ from
the nucleus (the vertical dashed lines in figure~\ref{pv12chmaj}).
Since non-circular motion, such as gas along a bar, generally has a
symmetric structure, we think that these structures in NGC 2146 may 
be produced by another mechanism.
Therefore, we averaged the $p-v$ diagram between $-4.8\arcsec$ and
$0\arcsec$ offset from the major axis (figure~\ref{sb2majmin}).
The outcome clearly shows a diffuse shell-like structure, indicated by the
red dashed curve labelled {\it SB2} in figure~\ref{sb2majmin}, only at
the western side of the galactic center.
Hence, this feature in the $p-v$ diagram can be explained
with in expanding molecular superbubble such as that in {\it SB1} 
mentioned above.

A close look at this superbubble shows two inner cavities,
indicated by red plus symbols in figure~\ref{sb2majmin}.
The velocities associated with {\it p1} and {\it p2}
are +20~km~s$^{-1}$ and -5~km~s$^{-1}$ relative to the systemic velocity,
respectively.
The diffuse structures around {\it p1} can be also seen in the $p-v$
diagrams (figure~\ref{pv12chmaj}) from the positions $-5\farcs0$ to
$+3\farcs0$ offset from the major axis, and those around {\it p2} appear
from $-5\farcs0$ to $0\farcs0$ relative to the major axis.
Furthermore, most of the $p-v$ diagrams in figure~\ref{pv12chmaj} show
that the molecular gas avoids the positions {\it p1} and {\it p2},
resulting in shell-like structures.
From these figures, we can infer two independent superbubbles at {\it p1}
and {\it p2}.
However, the signal-to-noise ratio difference between these two cavities
is small, only about $2\sigma$ difference, and therefore it is more
natural to consider these structures to be created by the same explosion
event.
In this case, the structure dividing {\it p1} and {\it p2} is just a
detail feature of one molecular bubble or possibly just noise in our data.
In addition, the positions of {\it p1} and {\it p2} lie away from the
rigid- or flat-rotation part where most of the molecular gas resides,
and hence, we consider the possibility that the molecular bubbles
originate from these points to be less likely.
Therefore, in the following, we will consider {\it SB2} as one
superbubble, which is $\sim$ 10$\arcsec$ offset from the galactic center
(figure~\ref{sb2majmin}).

From figures~\ref{m0rot43}, \ref{pv12chmaj}, and \ref{sb2majmin}, we
can estimate the main characteristics of {\it SB2}:
Its radius is estimated to be $\sim5\arcsec - 13\arcsec$ ($400-1000$~pc),
and the expansion velocity is measured as $\sim35\pm10$~km~s$^{-1}$.
The range of the radius depends again on where we assume the center of the
bubble to lie; the minimum radius corresponds to the edge of the disk, 
and the maximum to the plane of the disk, respectively.

\section{Search for Bubbles or Outflows by Velocity Model Subtraction}
\label{sect-model}

In the previous section, we have reported convincing arguments in favor of
the presence in NGC 2146 of two molecular bubbles and an outflow derived
using the integrated intensity maps and the $p-v$ diagrams.
In the following, we will focus on a search for bubbles and/or outflows
within the galactic disk of NGC 2146 using a `model subtraction' method.

The model we have used can be described as follows:
We first use the {\it AIPS} task {\it IRING} to compute mean
intensity within a range of azimuthal angles per a certain radius range of
a ring to obtain the velocity information.
We calculate the mean intensity within 10$\arcdeg$ of the major axis for
each redshifted and blueshifted velocity range (see figure~\ref{m0rot43}).
We then fit the $p-v$ diagram along the major axis with an
Elmegreen curve \citep{elm90}, parametrized as:
\begin{equation}
v = \frac{r}{(r^\alpha+r^{1-\beta})} \times \gamma,
\end{equation}
where $v$ is the rotational velocity in units of km~s$^{-1}$, $r$ is the
galactocentric radius in units of kpc, and the variables $\alpha$,
$\beta$, and $\gamma$ are the free parameters.
Figure~\ref{elm_fit} shows the data and the fit, and the results of the
fit are $\alpha = 0.30$, $\beta = 0.32$, and $\gamma = 256.39$.

To characterize the galactic kinematics of the central region of NGC 2146,
we compare the observed kinematics with the large-scale galactic rotation.
Figure~\ref{elm}a presents the $p-v$ diagram along the major axis
overplotted with the fitted rotation curve (from figure~\ref{elm_fit}).
We have further modeled the velocity map from the fitted rotation curve
under the assumption that the galactic disk is axisymmetric with an
inclination angle of 63$\arcdeg$ \citep{del99}.
Figure~\ref{elm}b shows this modeled velocity map overlaid with the
observed velocity map.
An excellent agreement between these two maps is seen.
We then subtract the modeled velocity map from our NMA velocity map,
and show the resultant residual velocity map in figure~\ref{elm}c
and the residual velocity along the major axis in figure~\ref{elm}d.
The residual velocities along the major axis are, on average, less than
$\pm$5~km~s$^{-1}$ (comparable to our velocity resolution) with maximum
residual velocities of only $\pm$10~km~s$^{-1}$.
In addition, the residual velocities over the entire galactic disk region
are never larger than about $\pm$20~km~s$^{-1}$.
There are some systematic residual velocities around $\pm4\arcsec$ from
the center with a negative residual velocity around the position of
$+4\arcsec$, and positive residual velocity around the position of
$-4\arcsec$.
Since these regions correspond to those where there are two rigid-rotation
components (see figure~\ref{pv_aspx}), a simple Elmegreen rotation curve
does not constitute an optimal fit.
These systematic residual velocities are therefore an artificial effect,
and should not be considered as an outflow feature.
From these results, we have concluded that the galactic kinematics 
within a few kpc of the central region of NGC 2146 can be described by 
normal galactic rotation.
We find no evidence of outflows with velocities greater than
20~km~s$^{-1}$ in the galactic disk.
Note that our model subtraction method does not detect any evidence of
{\it SB2}.
This is because our method uses an intensity-weighted mean velocity field
map, and therefore diffuse and weak components, such as {\it SB2}, cannot
be detected.

H\emissiontype{I} absorption line observations have revealed that the
central region of NGC 2146 is undisturbed and smooth, and can be
interpreted as a rotating disk \citep{tar04}, which is consistent with our
results.
The overall CO distributions in our channel maps are similar to those
shown in \citet{gre00}.
Some of the structures in the channel maps have been interpreted by
\citet{gre00} as outflow features (thick solid lines in their figures 6
and 8c).
Our velocity model fitting indicates that those structures
may be explained instead by normal galactic rotation.

\section{Properties of the Molecular Superbubbles and the Outflow}
\label{sect-prop}

In this section, we calculate the timescales, masses, and energies of the
molecular superbubbles and the outflow present in NGC 2146.
The molecular and total gas masses, in unit of M$_\odot$, can be calculated as
\begin{equation}
M_{\rm H_2} = 1.2 \times10^4 \times D^2 \times S_{\rm CO(1-0)}
                    \times \frac{X_{\rm CO}}{3.0 \times 10^{20}},
\label{eq_mh2}
\end{equation}
\begin{equation}
M_{\rm gas} = 1.36 \times M_{\rm H_2},
\label{eq_gas}
\end{equation}
where $D$ is the distance in units of Mpc, $S_{\rm CO(1-0)}$ is the CO
integrated intensity in units of Jy~km~s$^{-1}$, $X_{\rm CO}$ is the
CO-to-H$_{2}$ conversion factor in units of
cm$^{-2}$~(K~km~s$^{-1}$)$^{-1}$, and the factor 1.36 is to account for
the presence of elements other than hydrogen \citep{sak95}.
In the following calculations, we assumed the $X_{\rm CO}$ to be
$1.4\times10^{20}$~cm$^{-2}$~(K~km~s$^{-1}$)$^{-1}$, which is the same
value as that used for the molecular gas mass calculations in M82
\citep{mat00} with which NGC 2146 shares a number of similarities
(see section~\ref{sect-intro}).
We also assume that the energy of one supernova explosion is
$\sim1\times10^{51}$~erg \citep{ros98}.
All the derived values are summarized in table~\ref{tabbubl}.
Note that the total gas mass derived from our observation (with
$S_{\rm CO(1-0)}\sim1.8\times10^{3}$~Jy~km~s$^{-1}$; see
section~\ref{sect-overall}) is $\sim4.1\times10^{9}$~M${_\odot}$.

\subsection{Properties of Molecular Superbubbles}
\label{sect-sb}

First, we discuss the molecular superbubble {\it SB1}, which is
shown in figure~\ref{m0m1} and \ref{ab}.
If we assume that the expansion velocity stayed constant since the
beginning of the superbubble's expansion, we can derive its 
dynamical timescale, $t_{\rm SB1}$.
Using the radius of $\sim800-1200$~pc and the expansion velocity of
$50\pm10$~km~s$^{-1}$, we obtain the timescale
$t_{\rm SB1} = (800 - 1200~{\rm pc}) / (50\pm10~{\rm km~s}^{-1})
             \sim (1.3 - 2.9) \times 10^7~{\rm years}.$
The integrated intensity of {\it SB1} is 115~Jy~km~s$^{-1}$, so from
equations~(\ref{eq_mh2}) and (\ref{eq_gas}),
the gas mass of {\it SB1} is calculated to be
$\sim2.6\times10^8$~M$_\odot$.
The kinetic energy of the superbubble, $E_{\rm SB1}$, 
is therefore estimated to be 
$E_{\rm SB1} = \frac{1}{2} \times (2.6\times10^{8}~{\rm M_\odot})
                           \times (50\pm10~{\rm km~s^{-1}})^{2}
             \sim (4.1 - 9.3) \times 10^{54}~{\rm erg}.$
Considering that $\sim$10 -- 20 \% of the energy transmits from
supernova explosions into the surrounding ISM \citep{mcc87,wea77,lar74},
20,500 -- 93,000 supernova explosions are required to account for the
energy of {\it SB1}.

Second, we estimate the properties of the molecular superbubble {\it SB2},
which is shown in figures~\ref{m0rot43}, \ref{pv12chmaj}, and
\ref{sb2majmin}.
Again, with the same assumption of constant expansion velocity and using
the radius of $\sim400-1000$~pc (see section~\ref{sect-small}), 
we can derive the timescale, 
$t_{\rm SB2} = (400 - 1000~{\rm pc}) / (35\pm10~{\rm km\ s}^{-1})
             \sim (0.9 - 3.9) \times 10^7~{\rm years}.$
The integrated intensity of {\it SB2} measured from the $p-v$ diagram is
17.3~Jy~km~s$^{-1}$, so that the mass of {\it SB2} is calculated as
$\sim3.9\times10^7$~M${_\odot}$.
In this case, the energy of the bubble, $E_{\rm SB2}$, is estimated as
$E_{\rm SB2} = \frac{1}{2} \times (3.9\times10^{7}~{\rm M_\odot})
                           \times (35\pm10~{\rm km~s^{-1}})^{2}
             \sim (2.4 - 7.9) \times 10^{53}~{\rm erg}$.
After applying the efficiency of 10 -- 20\%,
the energy corresponds to 1,200 -- 7,900 supernova explosions.

The size, expansion timescales and velocities are similar for both
superbubbles, but the mass and therefore the energy derived for {\it SB1}
are about an order of magnitude larger than for {\it SB2}.
These results may suggest that {\it SB1} is related to a more active
(energetic) starburst than that related to {\it SB2}.

\subsection{Properties of the Molecular Outflow ``{\it OF1}''}
\label{sect-of}

Since the $p-v$ diagram shows a linear increase in velocity as a
function of the distance from the galactic disk for this molecular
outflow {\it OF1} (see section~\ref{sect-large}), it is natural to assume
that {\it OF1} is experiencing a constant acceleration.
Using a size and a terminal velocity for the outflow of 2~kpc and
200~km~s$^{-1}$, respectively, the acceleration, 
$a_{\rm OF1}$, and the timescale, 
$t_{\rm OF1}$, of the outflow can be calculated as
$a_{\rm OF1} = \frac{1}{2} \times (200~{\rm km~s^{-1}})^{2} / (2~{\rm kpc})$
		$\sim 3.2 \times 10^{-13}~{\rm km~s^{-2}}$
and
$t_{\rm OF1} = (200~{\rm km~s^{-1}}) / 
                   (3.2 \times 10^{-13}~{\rm km~s^{-1}})
             \sim 2.0 \times 10^{7}~{\rm years}$,
respectively.
On the other hand, it is also possible to assume that the initial
expansion velocity of the outflowing gas is different in each gas cloud.
Therefore faster (slower) gas is located far from (close to) the galactic
disk, resulting in the linear increase in velocity as a function of
the distance from the galactic disk in the $p-v$ diagram.
In this case, the timescale of {\it OF1} can be estimated from
the furthest distance of {\it OF1} and its velocity as,
$t'_{\rm OF1} = (2~{\rm kpc}) / (200~{\rm km~s^{-1}}) 
		\sim 1.0 \times 10^{7}~{\rm years}$. 
Both assumptions lead to the similar timescales.
The integrated intensity of {\it OF1} is estimated to be 150~Jy~km~s$^{-1}$,
leading to a total mass for the outflow of $3.4\times10^8$ M${_\odot}$
(see equations~\ref{eq_mh2} and \ref{eq_gas}).

According to the kinematics described above, the molecular outflow
velocity is not constant like in the case of the two molecular bubbles
{\it SB1} and {\it SB2}.
Hence, the outflow energy, $E_{\rm OF1}$ has to be calculated from each
channel map according to the relation
$E_{\rm OF1} = \sum_i\frac{1}{2}\ m_{i}v_{i}^2$,
where $m_{i}$ and $v_{i}$ are the molecular gas mass and the velocity of
{\it OF1} at each channel {\it i}.
The energy is calculated as
$E_{\rm OF1} \sim 3.0 \times 10^{55}~{\rm erg}$.
After applying the efficiency of 10 -- 20\%,
it corresponds to 150,000 -- 300,000 supernova explosions.

\section{Discussion}
\label{sect-dis}

\subsection{Soft X-ray Emission from the Large-Scale Molecular Bubble and
	Outflow}
\label{sect-xray}

NGC 2146 has a kpc-scale outflow from the galactic center detected at soft
X-ray wavelengths \citep{inu05,arm95}.
Here, we compare our \atom{C}{}{12}O(1-0) map with the soft X-ray image obtained
with the Chandra X-ray Observatory \citep{inu05}.
Figure~\ref{xray} shows the \atom{C}{}{12}O(1-0) integrated-intensity contour
map overlaid on the soft X-ray image.
The soft X-ray image shows strong extended emission toward the northeastern
part of the galaxy, and weak extended emission in the southwest.
Between these two diffuse emission regions, there is almost no emission
except for a number of strong point sources.
The latter location corresponds to the strong CO emitting region, namely
the galactic disk region, and therefore the lack of diffuse soft X-ray
emission can be explained by absorption from the molecular gas in the
galactic disk.
Because the stronger X-ray emission appears in the northeastern region of
the galactic disk rather than in the southwestern region, the most
probable picture is that the southwestern edge of the CO emission is at
the near side of the galactic disk and the northeastern one is at the far
side.
This scenario is indeed consistent with previous studies
\citep{arm95,gre00}.

Since the soft X-ray image traces the location of hot thermal plasma
in this galaxy \citep{arm95,del99,inu05}, the comparison between our
\atom{C}{}{12}O(1-0) map and the soft X-ray map gives the schematic
image of the relative location between the molecular gas and the hot
plasma.
In figure~\ref{xray}, the northeastern X-ray emission shows a conical
structure, emanating from the center of the galaxy and spreading outwards,
perpendicular to the major axis of the galaxy
(color-scale features enclosed by two red shorter dashed lines).
The molecular outflow, {\it OF1}, is located around the front edge of the
X-ray outflow.
This fact and the good spatial correspondence suggest that these two
outflows are physically connected.

We then compared the molecular gas outflow energy with that of the soft
X-ray outflow.
As reported in section~\ref{sect-of} (and in Table~\ref{tabbubl}), the
energy of {\it OF1} is calculated as
$E_{\rm OF1} \sim 3.0 \times 10^{55}$~erg.
On the other hand, the energy of the thermal plasma outflow is estimated
as $E_{\rm X} = 5.1 \times 10^{56}~f^{\frac{1}{2}}$~erg (with their value
re-scaled because of the different distance for NGC 2146 adopted by us;
\cite{inu05}), where $f$ is the volume filling factor for the X-ray
emitting gas.
If we adopt the standard value of 0.01 -- 0.1 for $f$ \citep{arm95},
the thermal plasma outflow energy can be calculated as
$(0.5-1.6)\times10^{56}$~erg.
The energy of the thermal plasma outflow is therefore comparable or an
order of magnitude larger than the molecular outflow energy, depending on
the volume filling factor for the X-ray emitting gas.
The thermal plasma would therefore have enough energy to push {\it OF1}
out from the galactic disk.

We also compared the pressure of the molecular gas in the outflow with
that of the soft X-ray outflow.
The molecular gas pressure can be measured by
$P_{\rm mol} \sim n({\rm H_{2}}) k_{\rm B} T$, where $n({\rm H_{2}})$
is the H$_{2}$ gas number density, $k_{\rm B}$ is the Boltzmann constant,
and $T$ is the molecular gas temperature.
Since our observations do not provide any density or temperature
information, here we assume standard values, namely a molecular gas
density and temperature traced by \atom{C}{}{12}O(1-0) of $10^{2-3}$~cm$^{-3}$
and 10 -- 100 K, respectively.
Thus the molecular gas pressure is estimated as
$\sim1.4\times10^{(-12\pm1)}$~dyne~cm$^{-2}$.
The thermal plasma outflow pressure is estimated as
$P_{\rm X} = (1.0 - 3.2)\times10^{-11}$~dyne~cm$^{-2}$ 
with the volume filling factor of 0.1 -- 0.01 \citep{inu05},
the pressure of the thermal plasma outflow is therefore comparable or two
orders of magnitude larger than that of the molecular outflow, depending
on the volume filling factor for the X-ray emitting gas and the density
and/or temperature of the molecular gas.
This result further supports the capability of the thermal plasma to
push {\it OF1} out from the galactic disk.
Since the energy and the pressure of the X-ray emission is larger than
that of the molecular outflow, it is natural to consider that the
expanding force from X-ray emission is still pushing the surrounding
molecular gas.

The distribution, kinematics, energies, and pressure of the molecular gas
and the hot thermal plasma outflows suggest that the hot thermal plasma
pushes the molecular gas outward, and produces the large-scale outflows.
The hot thermal plasma is caused by supernova explosions
in the central starburst region of NGC 2146 (e.g., \cite{arm95}) and,
very likely, the molecular gas outflow is also caused by the intense
starbursts at the center of NGC 2146 \citep{gre00}.
Furthermore, the arc-like feature of {\it OF1}, with the conical structure
of the hot thermal plasma outflow just behind {\it OF1}, suggests that the
molecular outflow could have had a bubble-like structure at the beginning
of the expansion, and then been blown out as an arc-like feature as
it evolved.
Indeed, the molecular superbubble with hot gas inside observed in the
central region of M82 \citep{mat05} could well represent an analogue of
{\it OF1} at an early stage.

Two-dimensional hydrodynamical simulations (e.g., \cite{tom88}) 
support the evolution of a molecular superbubble to an outflow.
One of their models simulated the evolution of a superbubble under
the ISM density of 100~cm$^{-3}$, which corresponds to that of
molecular gas seen in \atom{C}{}{12}O(1-0).
Supernova explosions at a constant rate for a long time produce a
vast amount of hot gas, and the hot gas accelerates and expands the
surrounding ISM gradually.
This model shows that the size of the molecular superbubble 
can expand to $\sim$2 kpc on a timescale of $\sim$10$^7$ years.
Since the supernova rate (SNR) of this model adopts that of M82
(0.1 yr$^{-1}$), and NGC 2146 has a similar SNR 
(0.15 yr$^{-1}$; \cite{tar00}), 
this model can be considered valid for NGC 2146.
Furthermore, the values of the size and the timescale from this model
are similar to our observations, and therefore our results are
consistent with the model.

At the center of the expanding molecular bubble {\it SB1}, diffuse soft
X-ray emission, indicative of hot thermal plasma, seems to be present.
Figure~\ref{xray} show two soft X-ray images with different spatial
resolutions:
Figure~\ref{xray}a is the higher spatial resolution image, where point
sources are clearly visible, while it is more difficult to see the
diffuse components.
Figure~\ref{xray}b is the lower spatial resolution image, where
diffuse components are more clearly seen.
Comparison of these two images can separate point sources and diffuse
sources, and indeed at the center of {\it SB1}, diffuse soft X-ray
emission is detected.
This situation is again similar to that observed in M82 and expected
from hydrodynamical simulations, suggesting that the hot gas is embedded

inside of the expanding molecular bubble {\it SB1}, pushing the
surrounding molecular gas outward.
A precise estimate of the energy of the soft X-ray component is, however,
difficult since, in this region, the soft X-ray emission is absorbed by
the foreground molecular gas and, therefore, any value derived for the
energy would be significantly underestimated.

\subsection{Configurations of Outflows and Bubbles}
\label{sect-conf}

Here we discuss the possible configuration of the outflows and bubbles in
the central $60\arcsec$ of NGC 2146 based on our CO and X-ray data, and
the previously published optical spectra.
A schematic diagram of the possible configuration of the molecular gas and
other components is shown in figure~\ref{conf} and, in the following
paragraphs, its details are discussed.

As shown in section~\ref{sect-model}, the molecular gas disk is well
modeled by the standard rigid- and flat-rotation curves with an
inclination of $63\arcdeg$.
Based on this simple regular galactic kinematics, it is safe to assume
that the galactic disk is flat and has no warp in the central region.
Comparison between the CO in the galactic disk and the X-ray data
indicates that the outflow is directed toward us in the northeastern side,
and the CO in the galactic disk in ``blocking'' the soft X-ray emission in
the southwestern side, making that the near side
(Sect.~\ref{sect-xray}).
Indeed, the long-slit optical spectra along the minor axis show that the
ionized gas is blueshifted to the northeast and redshifted to the
southwest, and the spectra taken in the central $2\arcsec$ show a
blue asymmetry in the emission lines, suggesting that the ionized gas
outflow is flowing toward us in the northeast, and the redshifted outflow
is blocked by the galactic disk around the galactic center \citep{arm95},
which is consistent with the configuration we propose.

The soft X-ray outflow shows a conical structure with a large opening
angle (see figure~\ref{xray}), and the outflowing direction is inclined
toward us.
From the blue asymmetry in the optical spectra near the galactic center,
the line-of-sight ionized gas outflow velocity toward us is estimated as
$\sim300-400$~km~s$^{-1}$ \citep{arm95}.
The redshifted side, on the other hand, shows only $\sim100$~km~s$^{-1}$
(estimated from figure~11 of \cite{arm95}), suggesting that the far
side of the conical structure of the outflow is nearly perpendicular to
the line-of-sight.
{\it OF1} is dominated by the redshifted gas, suggesting that most of the
molecular gas is located at the far side of the conical structure.
The line width of ionized gas increases with distance 
to the galactic center \citep{arm95},
suggesting an acceleration as the gas moves away, 
consistent with the kinematics of {\it OF1}.

{\it SB1} is located in the southwestern part of the galaxy, indicating
that {\it SB1} is at the farside of the galactic disk.
The radius of the molecular gas disk is about 2~kpc (about $25\arcsec$)
with an inclination of $63\arcdeg$, while the central velocity of
{\it SB1} is somewhat blueshifted and the radius of {\it SB1} is about
$\sim1$~kpc.
Therefore, in order to be able to partially see {\it SB1},
this has to be closer to us than the galactic center.

Summing up all the aforementioned information, we made a schematic
diagram of the possible configuration of the molecular gas and the hot
gas outflows in figure~\ref{conf}.
In the figure, the molecular gas is shown in red, X-ray emission in
yellow, and optical emission line in light blue.
The emission obscured by the molecular gas (and possibly dust that is
mixed with the molecular gas) in front is shaded.
The overall configuration of this schematic diagram is globally consistent
with that presented in \citet{gre00} but, by using our new observations,
we have been able to describe the different structures and components
presented in the inner region of NGC 2146 in greater detail.

In sections~\ref{sect-sb} and \ref{sect-of},
we derive the properties of {\it SB1} and {\it OF1}.
Since the timescale of {\it SB1} corresponds roughly to that of {\it OF1},
it is possible that {\it SB1} and {\it OF1} originated from
one starburst event.
The difference in the structure ({\it SB1} is still a bubble while
{\it OF1} is already outflowing) can be explained by the
possible location where the starburst may have started up;
if a starburst occured a bit above the galactic disk plane, 
an expansion upward is less frustrated than downward due to a less dense ISM.
Since a bubble expands faster in less dense conditions \citep{tom88}, 
an expansion in a lower density ISM leads to faster bubble evolution, 
and the bubble becomes an outflow earlier than that in a dense region.
However, the possible central locations for {\it SB1} and {\it OF1}
are different (figure~\ref{m0m1}; see also figure~\ref{conf}), it is
difficult to say that {\it SB1} and {\it OF1} have the same explosion
center.
Thus it is safer to conclude that {\it SB1} and {\it OF1}
may have formed in the same explosion period,
but at different locations.

\section{Summary}
\label{sect-sum}

We observed the nearby nearly edge-on starburst galaxy NGC 2146 with the NMA
in the \atom{C}{}{12}O(1-0) line.
We performed a deep integration and successfully imaged a diffuse
molecular outflow ({\it OF1}) and two superbubbles ({\it SB1} and
{\it SB2}), expanding from the galactic disk region.

{\it OF1} is located toward the northeastern side of the galactic disk and
shows a linear increase in velocity with the distance from the galactic
disk.
It shows an arc-like feature, and is located at the front edge of the soft
X-ray outflow.
The energy of {\it OF1} is comparable to or an order of magnitude lower
than that of the soft X-ray outflow.
The pressure of {\it OF1} is about one or two orders of magnitude
lower than that of the soft X-ray outflow.
Therefore these results suggest that {\it OF1} is pushed outward by the
energetic plasma outflow from the galactic disk.

{\it SB1} is located at the southwestern side of the galactic disk and
shows an expanding motion.
The soft X-ray image shows diffuse and weak emission at the center of
{\it SB1}.
Since {\it SB1} is located on the far side of the galactic disk from us,
this weakness is likely due to the soft X-ray emission being highly
absorbed by the galactic disk.
It is not clear whether the soft X-ray emission can be the energy
source of the expansion of {\it SB1}.
The lifetime of {\it SB1} and {\it OF1} are similar, and hence, we think
that these structures have been formed during the same explosion period.
{\it SB2} is clearly visible in the position-velocity diagram, and has
less mass than {\it SB1}, and is therefore less energetic than
{\it SB1}.
Possibly {\it SB2} has been caused by a less energetic starburst than
the one that produced {\it OF1} or {\it SB1}.

The galactic disk seems to have two rigid-rotating components and one
flat-rotating component.
The inner rigid-rotating component has a steeper velocity gradient than
the outer rigid-rotating component, and is $\sim300$~pc in radius,
indicating that there is a fast rotating nuclear disk in this galaxy.
Model fitting to the velocity field maps shows that the central region
of the galactic disk of NGC 2146 can be explained by regular galactic
rotation, and there is no clear evidence for molecular gas tracing an
outflow inside the galactic disk.

\bigskip

We appreciate to Paul T. P. Ho, You-Hua Chu, and the anonymous referee
for very useful comments.
We also grateful to the NRO staff for the operation and improvement of
the NMA.
This work is supported by the National Science Council (NSC) of
Taiwan, NSC 96-2112-M-001-009 and NSC 97-2112-M-001-021.

\begin{figure*}
\begin{center}
\FigureFile(160mm,160mm){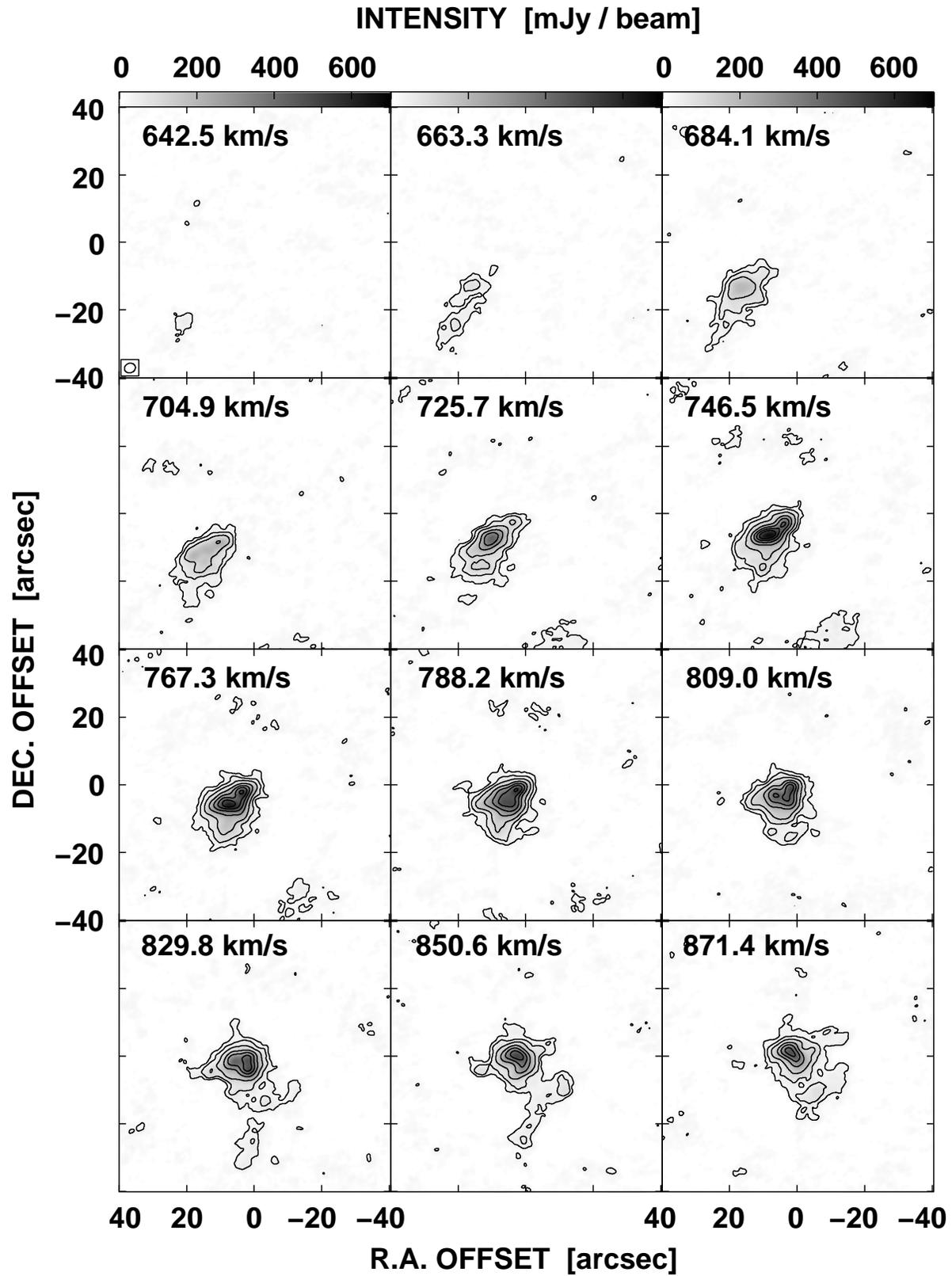}
\end{center}
\caption{
\atom{C}{}{12}O(1-0) channel maps of NGC 2146.
The central position and the synthesized beam size are the same as in
figure~\ref{m0m1}, and the beam is shown in the lower left corner of panel 1.
The contour levels are 5, 10, 20, 40, 60, 80, and 100$\sigma$,
where 1$\sigma$ is 5.7~mJy~beam$^{-1}$.
The LSR velocity is showing at the top left corner of each channel map.
}
\label{ch0112}
\end{figure*}
\setcounter{figure}{1}
\begin{figure*}
\begin{center}
\FigureFile(160mm,160mm){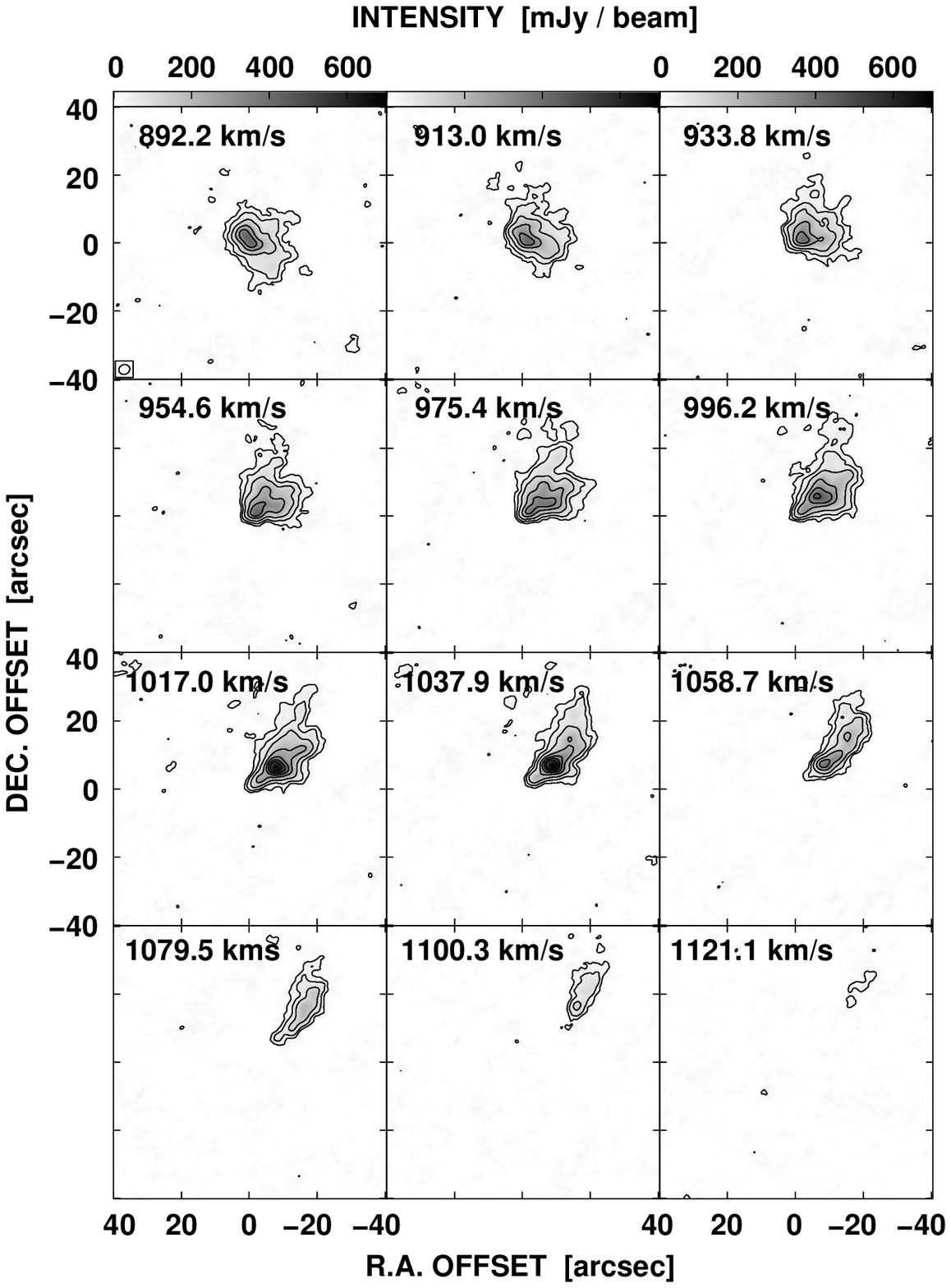}
\end{center}
\caption{
Continued.
}
\label{ch1324}
\end{figure*}

\begin{figure}
\begin{center}
\FigureFile(80mm,80mm){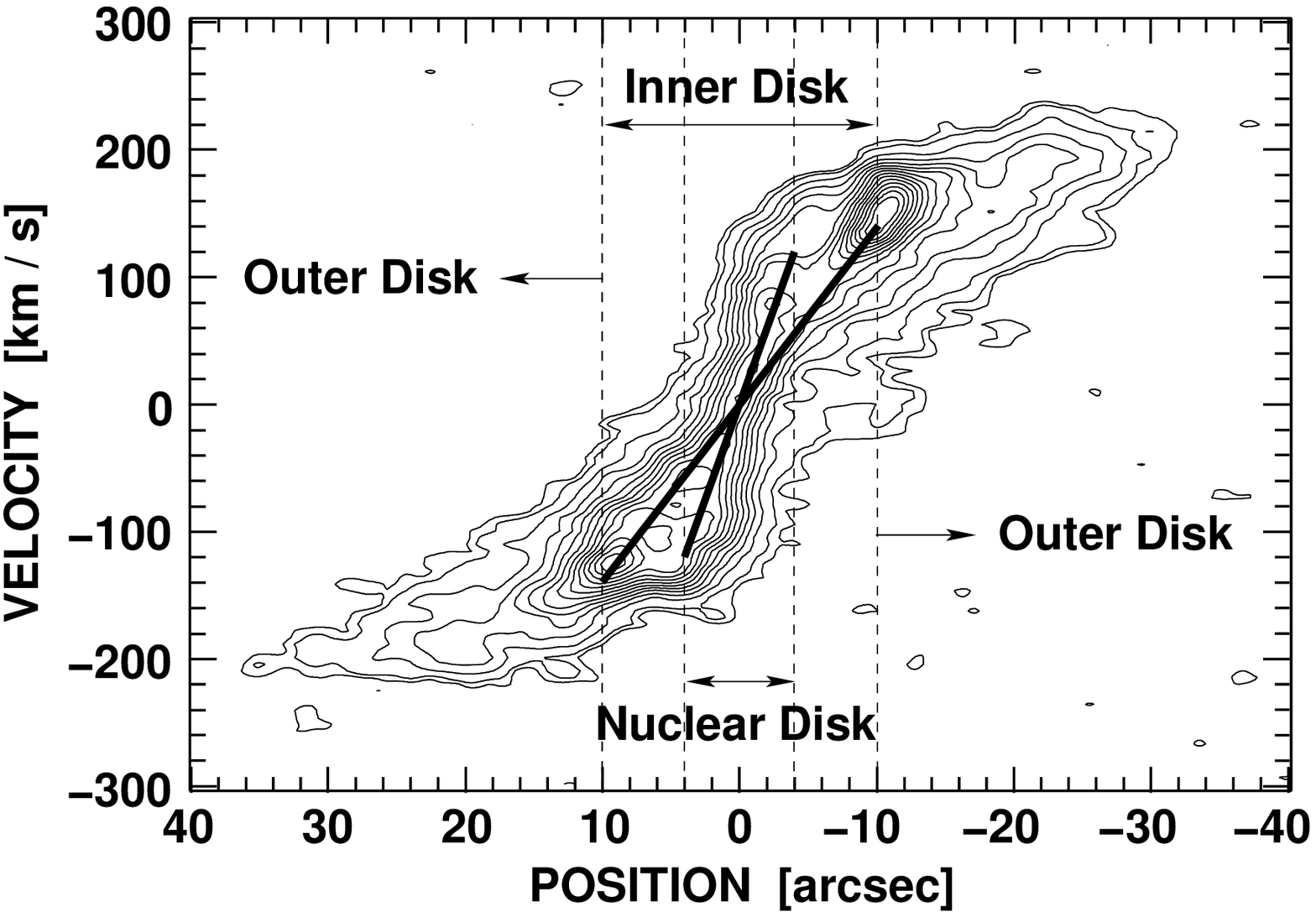}
\end{center}
\caption{
The $p-v$ diagram of NGC 2146 along the major axis.
The central position corresponds to R.A.\ = \timeform{6h18m37s.6},
and the central LSR velocity corresponds to 900~km~s$^{-1}$.
The contour levels are 3, 5, 10, 15, 20, $\dots$, 65, and 70$\sigma$,
where 1$\sigma$ is 0.01~Jy~beam$^{-1}$.
The vertical dashed lines between $\sim\pm4\arcsec$ indicate the region of
the nuclear disk, and those between $\sim\pm10\arcsec$ indicate the region
of the inner disk, which are defined by the two rigid-rotating components
shown in the thick solid lines.
The outer regions, where kinematics are dominated by flat-rotation, are
the outer disk.
}
\label{pv_aspx}
\end{figure}

\begin{figure}
\begin{center}
\FigureFile(80mm,80mm){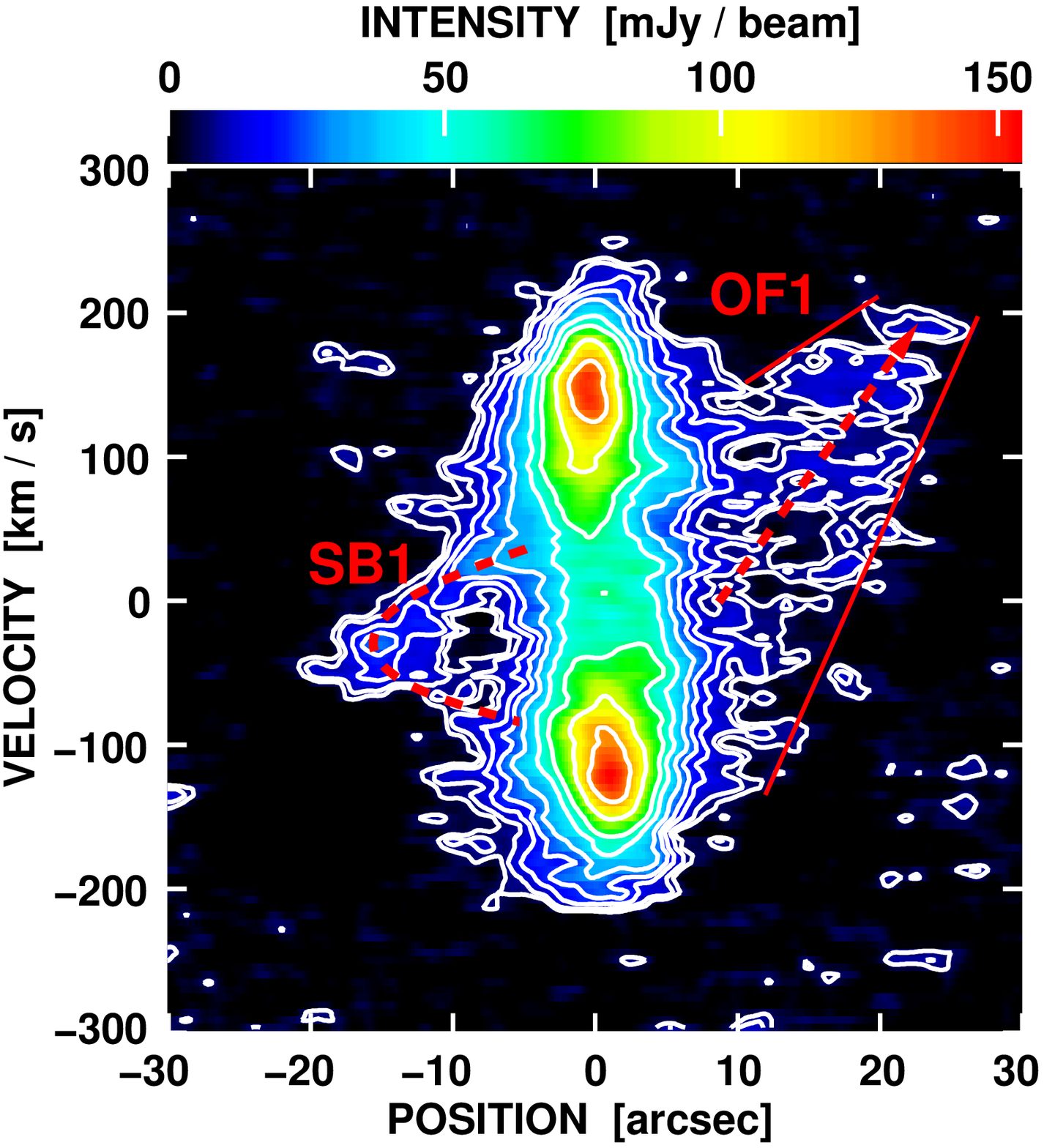}
\end{center}
\caption{
Major-axis-integrated $p-v$ diagram along the minor axis.
This $p-v$ diagram is obtained by integrating the data along the major
axis of the range between the position $+22\arcsec$ and $-28\arcsec$,
which is between two dotted vertical lines in figure~\ref{m0rot43}.
The position of zero corresponds to the major axis, and the zero velocity
corresponds to the systemic velocity of 900~km~s$^{-1}$.
The contour levels are 2, 3, 5, 7, 10, 15, 20, 30, and 40$\sigma$,
where 1$\sigma$ is 3.2~mJy~beam$^{-1}$.
Here {\it SB1} and {\it OF1}, which are displayed in red dashed lines,
indicate the positions of an expanding bubble and an outflow, which are
shown in figure~\ref{m0m1}a (see section~\ref{sect-large} for details).
Two solid red lines indicate the range of the outflow.
}
\label{ab}
\end{figure}

\begin{figure}
\begin{center}
\FigureFile(80mm,80mm){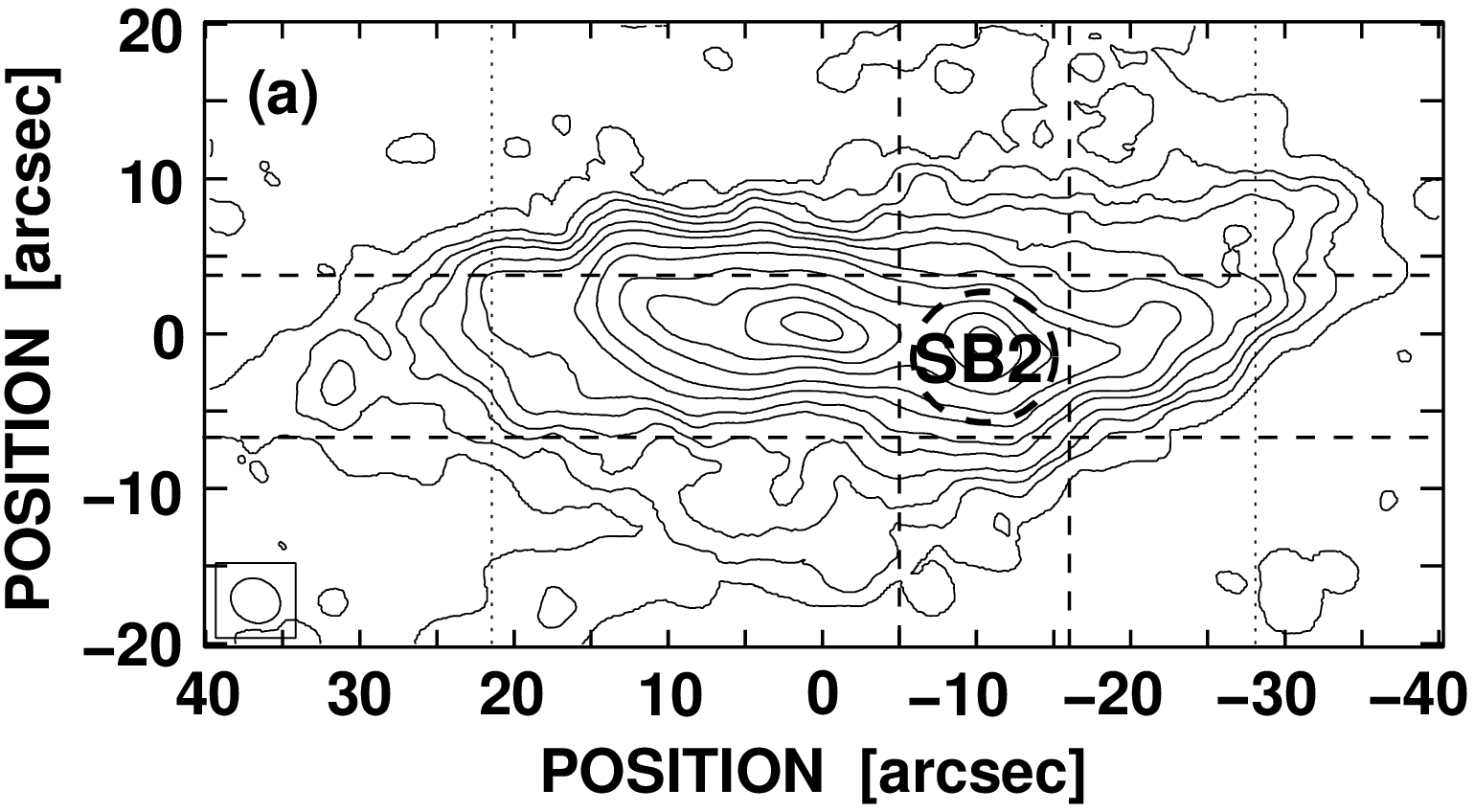}
\FigureFile(80mm,80mm){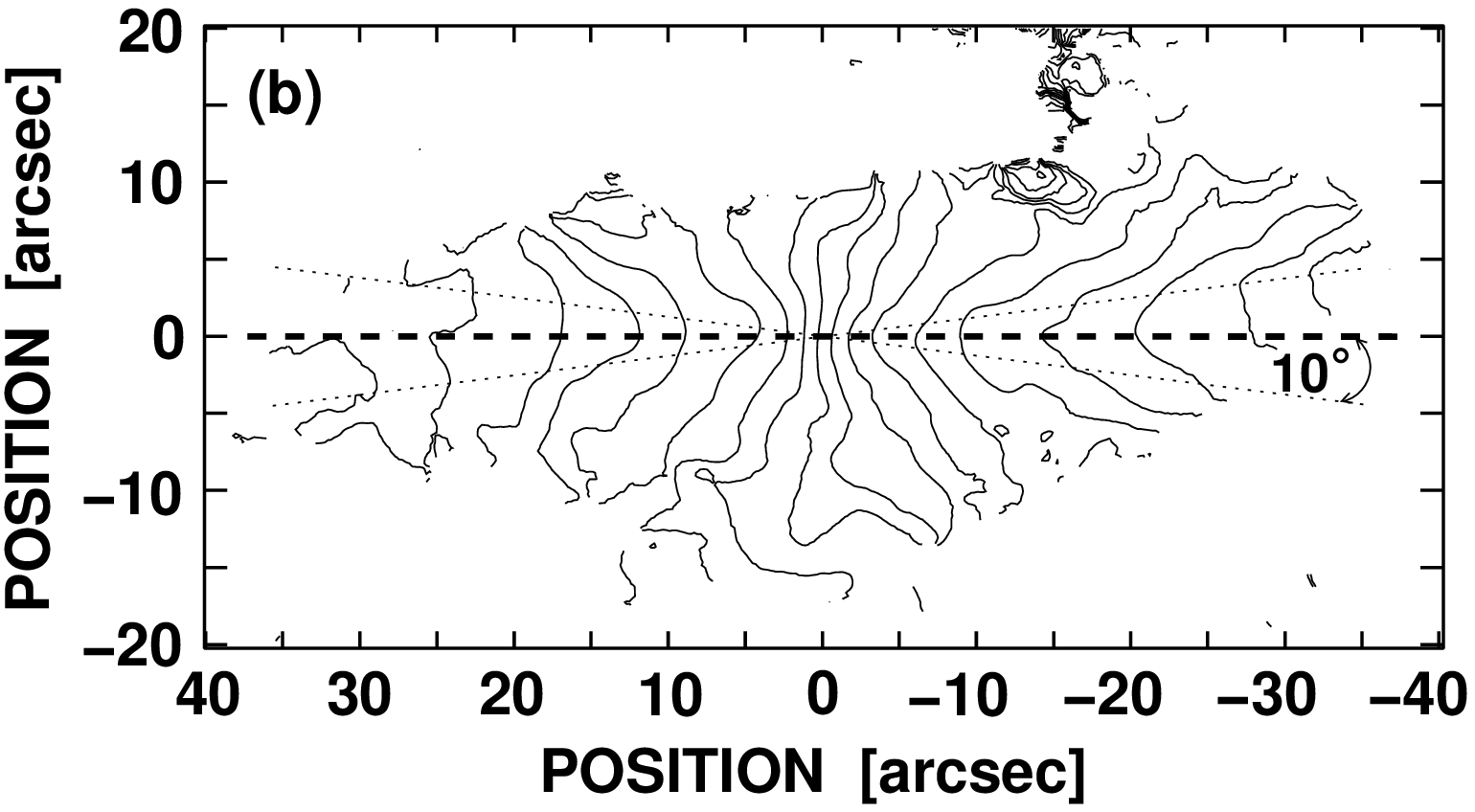}
\end{center}
\caption{
(a) The integrated intensity map after a clockwise rotation of
    $137\arcdeg$.
    The contour levels are the same as in figure~\ref{m0m1}a.
    Th two horizontal dashed lines, offset by $-7\arcsec$ and $+4\arcsec$ 
    from the major axis, specify the range of major axis $p-v$ diagrams
    shown in figure~\ref{pv12chmaj}.
    The two vertical dashed lines, $-5\arcsec$ and $-16\arcsec$ 
    from the minor axis, specify the range of the small molecular
    bubble {\it SB2}, and are the same as the two vertical dashed lines
    in figures~\ref{pv12chmaj} and \ref{sb2majmin}.
    The thick dashed circle therefore indicates the region of {\it SB2}.
    The two vertical dotted lines indicate the integrated region of the
	$p-v$ diagram in figure~\ref{ab}, which is between $+22\arcsec$ and
	$-28\arcsec$ along the minor axis.
(b) The velocity field map after a clockwise rotation of $137\arcdeg$.
    The contour levels are the same as in figure~\ref{m0m1}b.
    The thick horizontal dashed line indicates the major axis.
    The two dotted lines specify the region (within $\pm10\arcdeg$ from the
    major axis) where the data points for the rotation curve shown in
    figure~\ref{elm_fit} were collected.
}
\label{m0rot43}
\end{figure}

\begin{figure*}
\begin{center}
\FigureFile(150mm,150mm){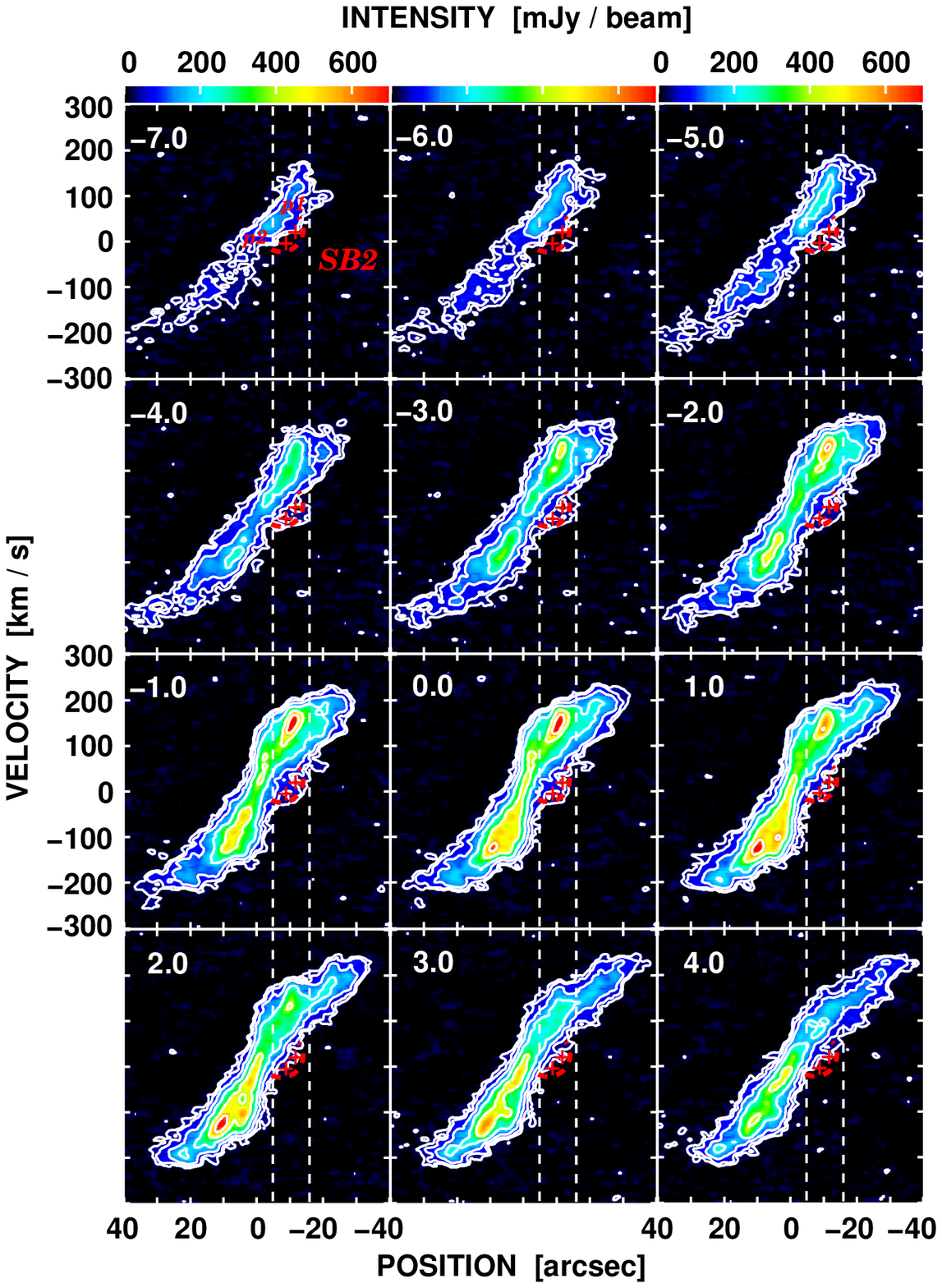}
\end{center}
\caption{
The $p-v$ diagrams along the major axis with various offsets.
Each $p-v$ diagram is averaged every 2$\arcsec$ with an interval of
1$\arcsec$.
The contour levels are 3, 5, 10, 20, 40, and 60$\sigma$,
where 1$\sigma$ is 10~mJy~beam$^{-1}$.
The number in the top-left corner of each $p-v$ diagram shows the position
offset in units of arcseconds from the major axis in figure~\ref{m0rot43},
which is offset by $-7\arcsec$ to $+4\arcsec$ from the major axis
(see also figure~\ref{m0rot43}).
The red plus symbols, {\it p1} and {\it p2}, mark the possible centers of
molecular bubbles.
The velocity of the red plus symbols {\it p1} and {\it p2} are 
920 and 895~km~s$^{-1}$, respectively.
The position of two vertical dashed lines are the same as in
figure~\ref{m0rot43}, namely, $-5\arcsec$ and $-16\arcsec$ offset from
the minor axis.
}
\label{pv12chmaj}
\end{figure*}

\begin{figure}
\begin{center}
\FigureFile(80mm,80mm){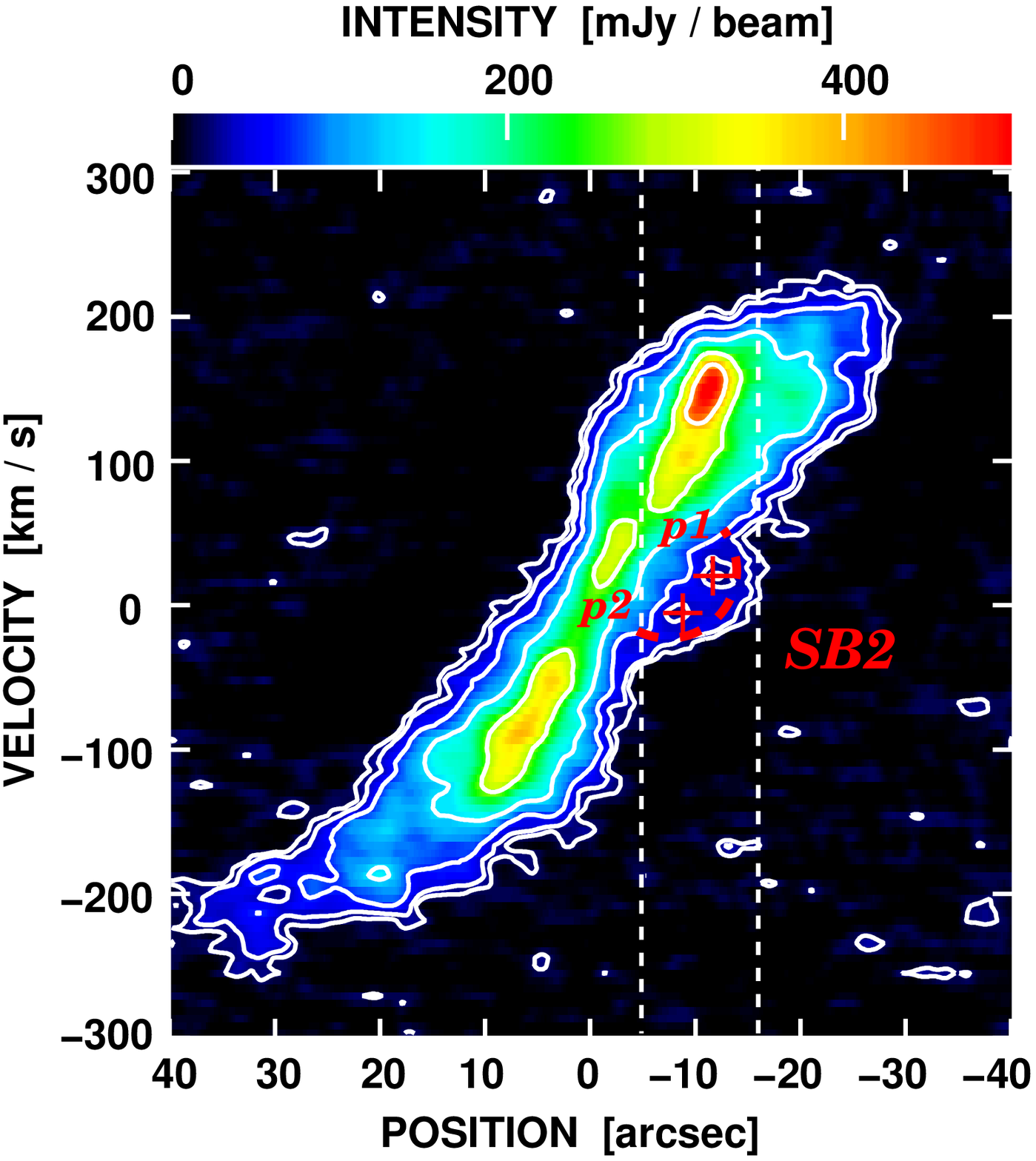}
\end{center}
\caption{
The averaged $p-v$ diagram of NGC 2146 and diffuse molecular bubbles
{\it SB2}.
The $p-v$ diagram was made by averaging between $-4\farcs8$ and $0\farcs0$
offset from the major axis.
The symbols {\it p1} and {\it p2} and the two vertical dashed lines are
the same as in figure~\ref{pv12chmaj}.
The red dashed curve indicates the possible position of the molecular
bubble {\it SB2}.
The contour levels are 3, 5, 10, 20, 40, and 60$\sigma$,
where 1$\sigma$ is 6.8~mJy~beam$^{-1}$.
}
\label{sb2majmin}
\end{figure}

\begin{figure}
\begin{center}
\FigureFile(80mm,80mm){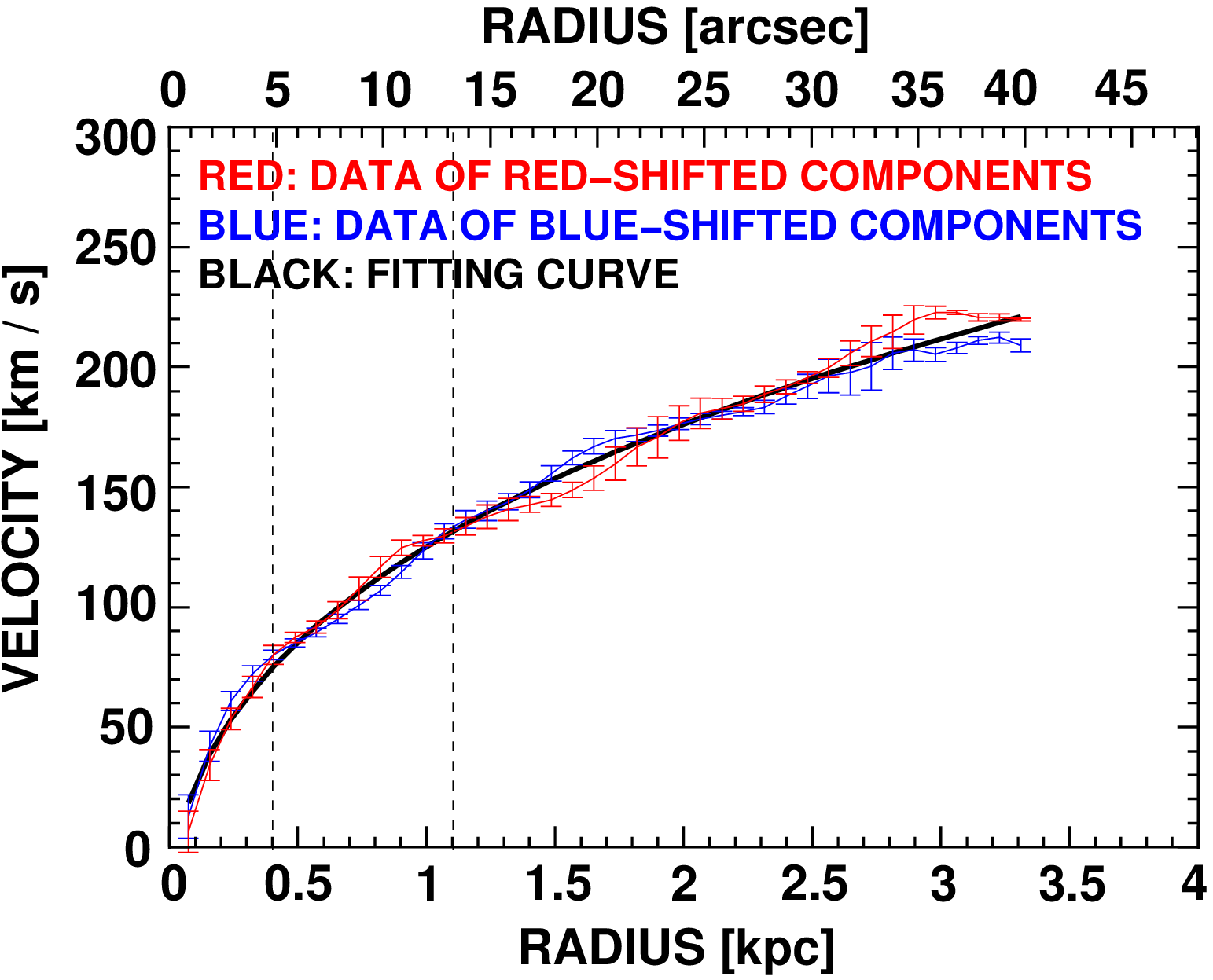}
\end{center}
\caption{
Observed and fitted rotation curve along the major axis.
The horizontal axis is the distance to the galactic center.
The units in the lower axis are in kiloparsecs, 
and in the upper axis, arcseconds.
The red and blue data are the redshifted and blueshifted components,
repectively.
The Elmegreen rotation curve is used for the fitting, and the black thick
solid curve is the fitted rotation curve from both redshifted and
blueshifted data.
The vertical dashed lines in the radii of 4$\arcsec$ and 10$\arcsec$
indicate the sizes of the nuclear and the inner disks, respectively.
}
\label{elm_fit}
\end{figure}

\begin{figure*}
\begin{center}
\FigureFile(80mm,80mm){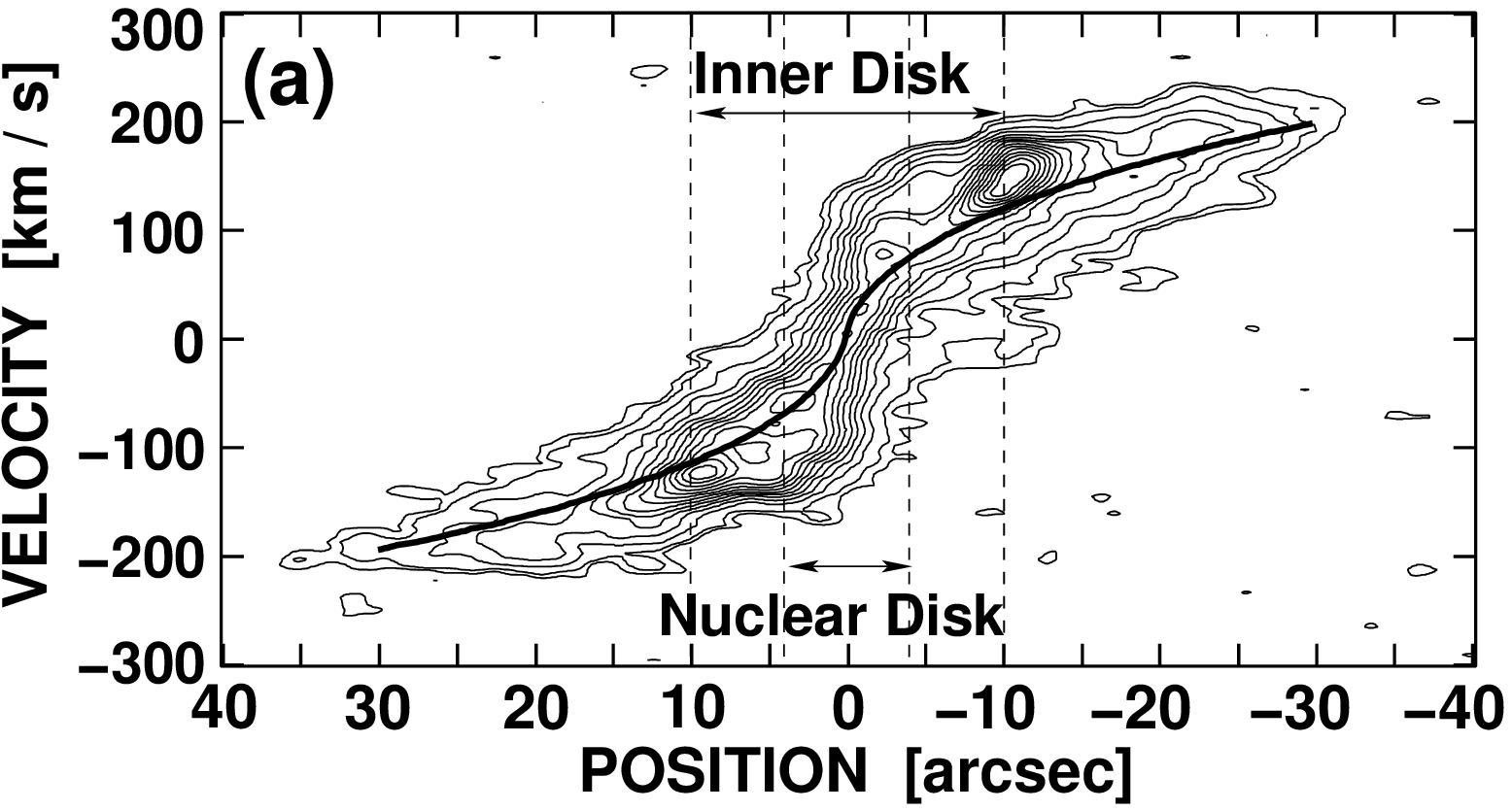}
\FigureFile(80mm,80mm){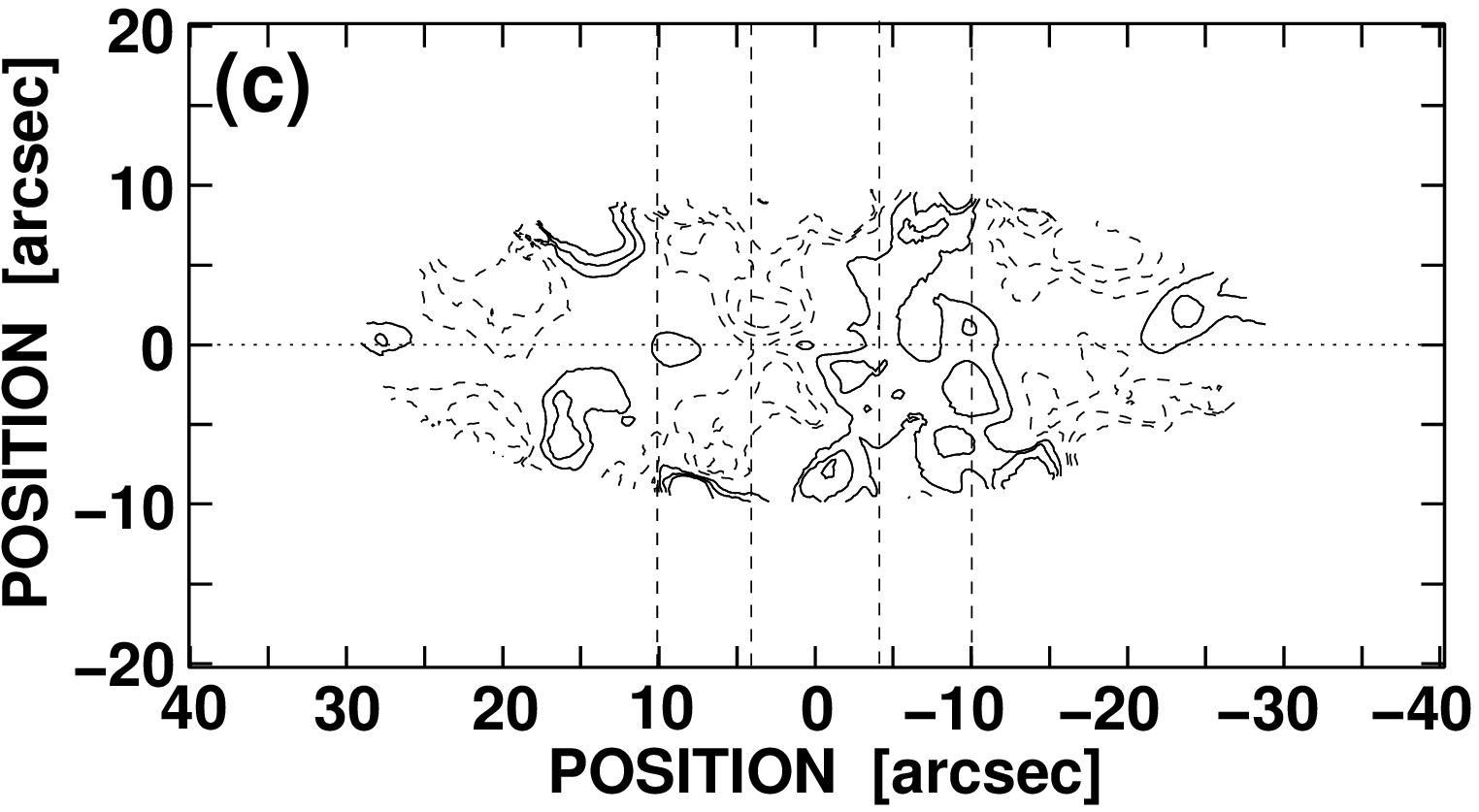}
\FigureFile(80mm,80mm){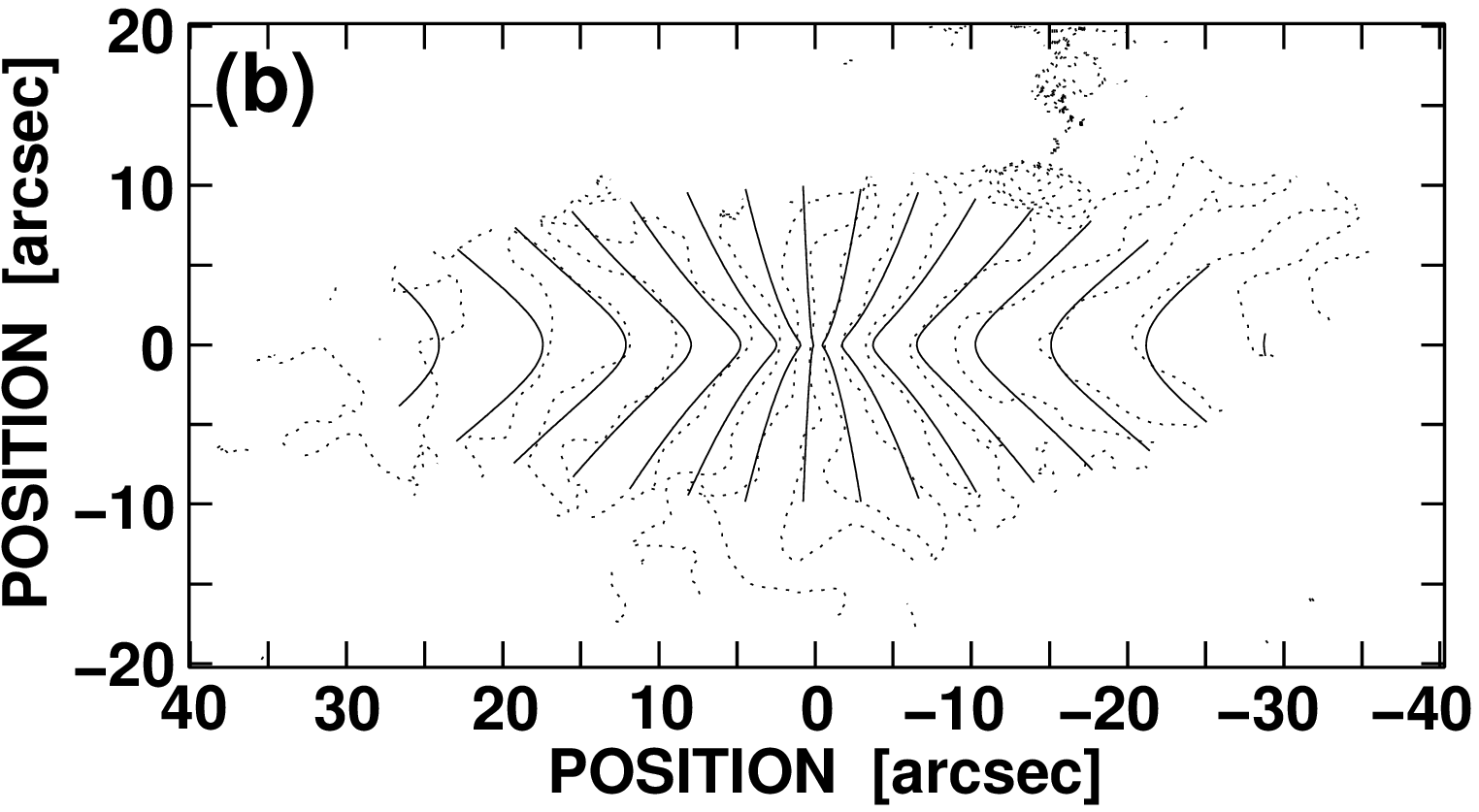}
\FigureFile(80mm,80mm){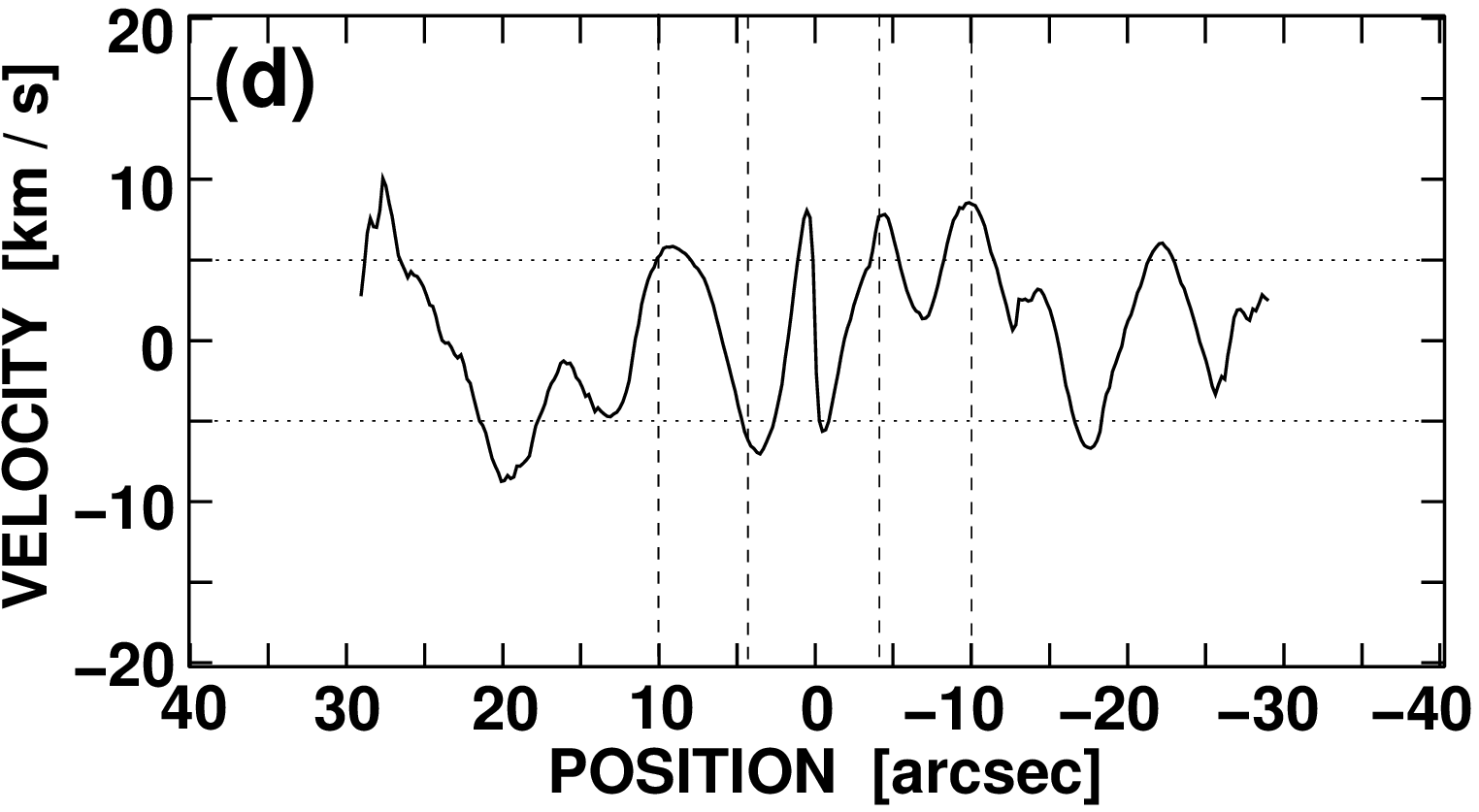}
\end{center}
\caption{
(a) The $p-v$ diagram with the fitted rotation curve.
    The contours are the $p-v$ diagram along the major axis, which is the
    same as in figure~\ref{pv_aspx}.
    The black thick solid curve is the fitted rotation curve in
    figure~\ref{elm_fit}.
    The vertical dashed lines show the range of the nuclear and inner
    disks.
(b) A modeled velocity field map overlaid on the observed velocity map.
    The solid lines display the modeled velocity map, which is created
    from the fitted rotation curve in figure~\ref{elm}a.
    The contour levels, from left to right, are 720, 745, 770, $\dots$,
    1095~km~s$^{-1}$, increasing with 25~km~s$^{-1}$.
    The dotted lines are the velocity field map, which is the same as in
    figure~\ref{m0m1}b.
(c) A residual velocity map.
    The residual velocity map is the differential values between
    the observed and the modeled velocity field maps.
    The solid contours are 5, 10, and 15~km~s$^{-1}$, and the dashed
    contours are -5, -10, and -15~km~s$^{-1}$.
(d) A residual velocity along the major axis.
    The horizontal dotted lines indicates the velocity within
    $\pm5.2$~km~s$^{-1}$, which is our velocity resolution.
	See Sect.~\ref{sect-model} for more details.
}
\label{elm}
\end{figure*}

\begin{figure*}
\begin{center}
\FigureFile(80mm,80mm){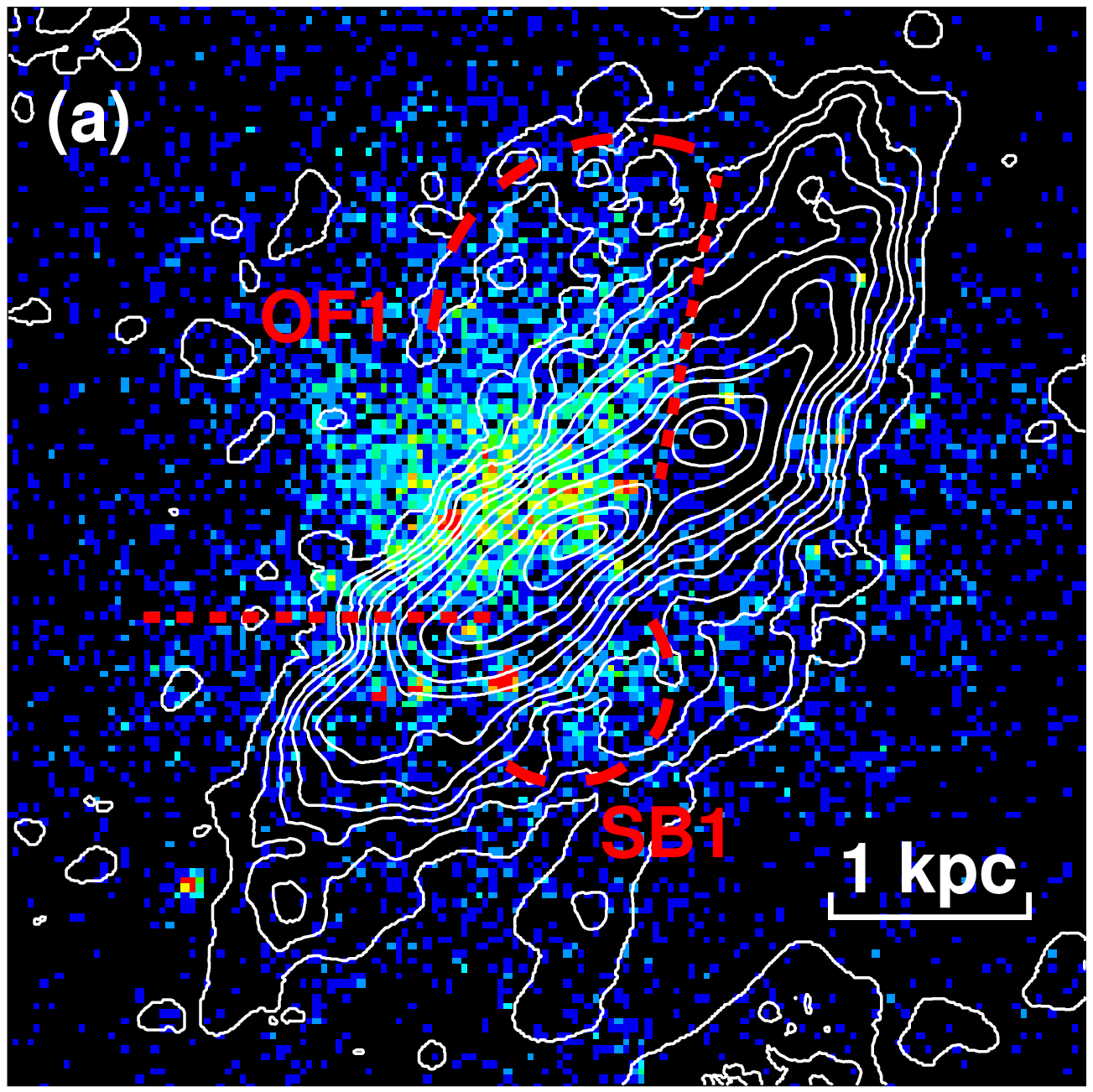}
\FigureFile(80mm,80mm){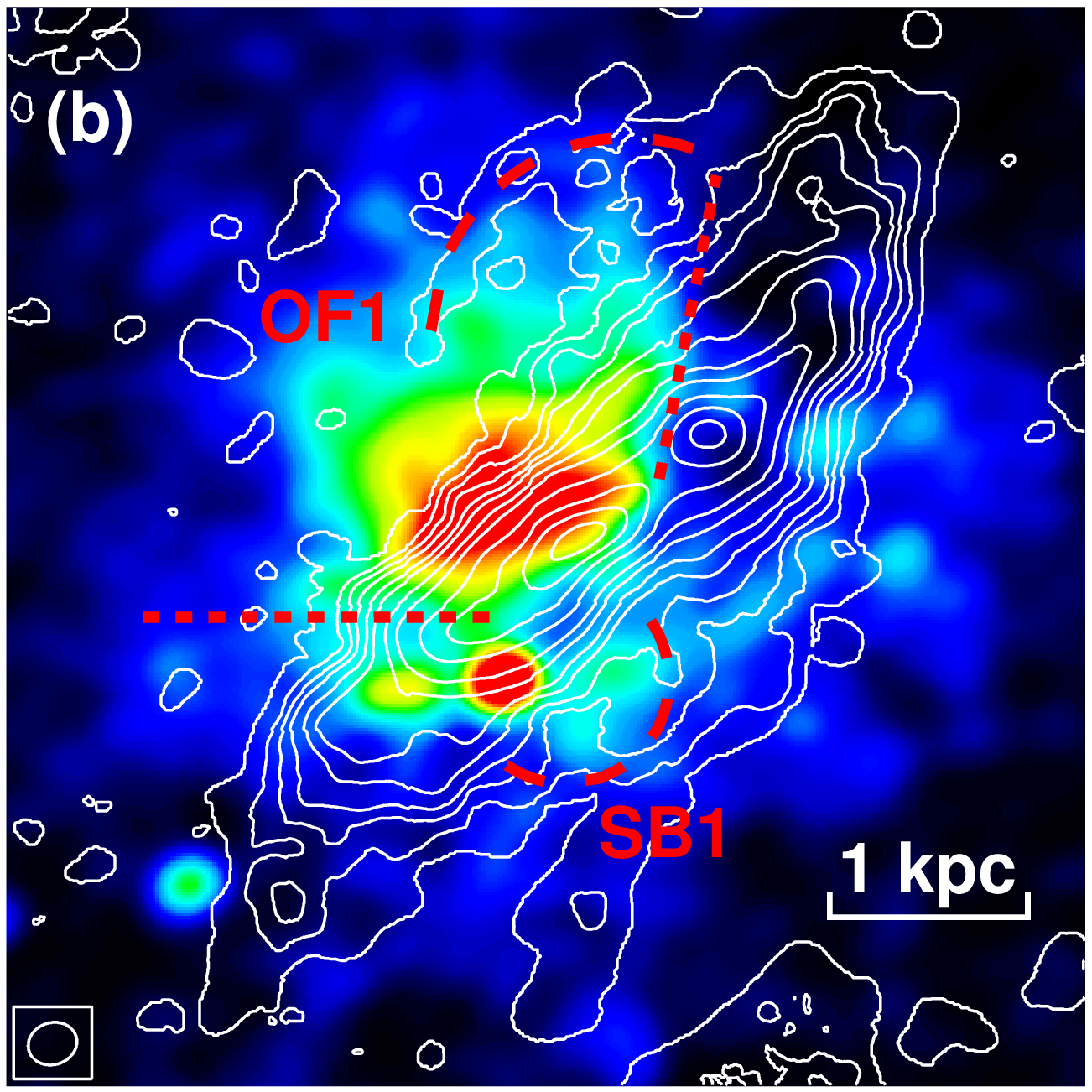}
\end{center}
\caption{
NMA \atom{C}{}{12}O(1-0) integrated intensity contour map overlaid on the
Chandra soft X-ray image \citep{inu05} in color-scale.
The contour levels of the \atom{C}{}{12}O(1-0) map and the red long dashed lines marked
as {\it SB1} and {\it OF1} are the same as in figure~\ref{m0m1}a.
The X-ray outflow is collimated as shown between two red short dashed
lines.
Some weaker X-ray emission is concentrated inside {\it SB1}.
(a) Without smoothing the soft X-ray image.
(b) Smoothed to a beam size of $2\farcs8 \times 3\farcs4$ 
(the same with that of the NMA image).
}
\label{xray}
\end{figure*}

\begin{figure}
\begin{center}
\FigureFile(80mm,80mm){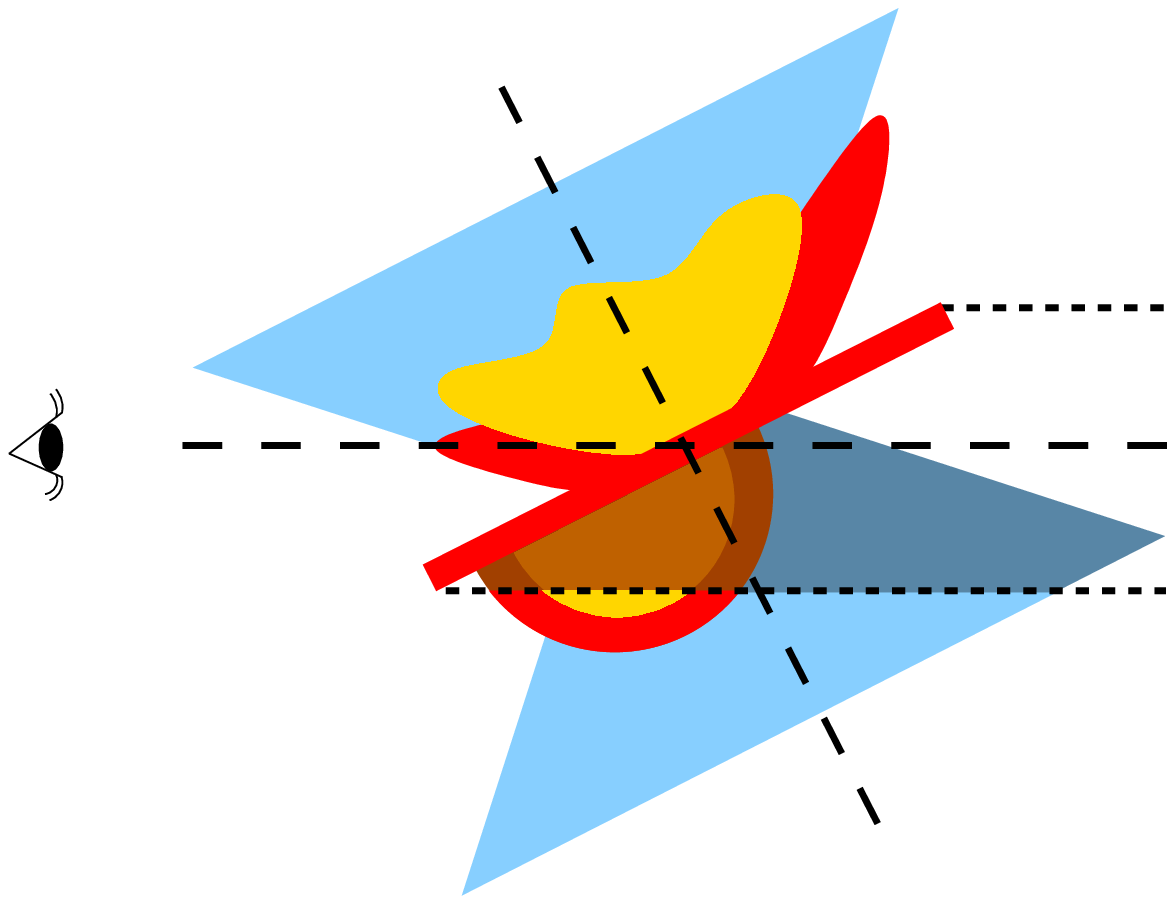}
\end{center}
\caption{
The schematic diagram of outflows and superbubbles in NGC 2146
along the line-of-sight.
The observer is on the left.
The red colored region indicates the CO distribution,
the yellow colored region indicates the soft X-ray emission distribution,
and the blue colored region indicates optical emission line distribution.
The gas behind the galactic disk is shaded to show it is absorbed by
the galactic disk.
}
\label{conf}
\end{figure}

\begin{table*}
\begin{center}
\caption{\bf The properties of the two Molecular Superbubbles, {\it SB1}
	and {\it SB2}, and of the Outflow {\it OF1}}
\label{tabbubl}
\begin{tabular}{lccccc}
\hline
Component	& Radius
		& Velocity
		& Timescale
		& Mass
		& Energy
\\
	& [pc]
	& [km s$^{-1}$]
	& [10$^7$ yr]
	& [10$^8$ M$_\odot$]
	& [10$^{54}$ erg]
\\
\hline
{\it SB1}	& 800 -- 1200
		& 50$\pm$10
		& 1.3 -- 2.9
		& 2.6
		& 4.1 -- 9.3
\\
{\it SB2}	& 400 -- 1000
		& 35$\pm$10
		& 0.9 -- 3.9
		& 0.39
		& 0.24 -- 0.79
\\
{\it OF1}	& 2000
		& 0 -- 200
		& 1.0 -- 2.0 
		& 3.4
		& 30
\\
\hline
\end{tabular}
\end{center}
\end{table*}


\begin{thebibliography}{}
\bibitem[Armus et al.(1995)]{arm95} Armus, L., Heckman, T. M.,
	Weaver, K. A., \& Lehnert, M. D.  1995, \apj, 445, 666
\bibitem[Benvenuti et al.(1975)]{ben75} Benvenuti, P., Capaccioli, M.,
	\& D'Odorico, S.  1975, \aap, 41, 91
\bibitem[Dahlem et al.(1998)]{dah98} Dahlem, M., Weaver, K. A.,
	\& Heckman, T. M.  1998, \apjs, 118, 401
\bibitem[Della Ceca et al.(1999)]{del99} Della Ceca, R.,
	Griffiths, R. E., Heckman, T. M., Lehnert, M. D.,
	\& Weaver, K. A.  1999, \apj, 514, 772
\bibitem[Deul \& den Hartog(1990)]{deu90} Deul, E. R.,
	\& den Hartog, R. H.  1990, \aap, 229, 362
\bibitem[Dumke et al.(2001)]{dum01} Dumke, M., Nieten, Ch., Thuma, G.,
	Wielebinski, R., \& Walsh, W.  2001, \aap, 373, 853
\bibitem[Elmegreen \& Elmegreen(1990)]{elm90} Elmegreen B. G.
	\& Elmegreen D. M.  1990, \apj, 355, 52
\bibitem[Fisher \& Tully(1976)]{fis76} Fisher, J. R.,
	\& Tully, R. B.  1976, \aap, 53, 397
\bibitem[Greve et al.(2006)]{gre06} Greve, A., Neininger, N., Sievers, A.,
	\& Tarchi, A.  2006, \aap, 459, 441
\bibitem[Greve et al.(2000)]{gre00} Greve, A., Neininger, N., Tarchi, A.,
	\& Sievers, A.  2000, \aap, 364, 409
\bibitem[Handa et al.(1992)]{han92} Handa, T., Sofue, Y., Ikeuchi, S.,
	Kawabe, R., \& Ishizuki, S.  1992, \pasj, 44, L227
\bibitem[Hutchings et al.(1990)]{hut90} Hutchings, J. B., Neff, S. G.,
	Stanford, S. A., Lo, E., \& Unger S. W.  1990, \aj, 100, 60
\bibitem[Inui et al.(2005)]{inu05} Inui, T., Matsumoto, H., Tsuru, T. G.,
	Koyama, K., Matsushita, S., Peck, A. B., \& Tarchi, A.  2005, \pasj,
	57,135
\bibitem[Irwin \& Sofue(1996)]{irw96} Irwin, J. A. \& Sofue, Y.  1996,
	\apj, 464, 738
\bibitem[Jackson \& Ho(1988)]{jac88} Jackson, J. M.,
	\& Ho, P. T. P.  1988, \apjl, 324, L5
\bibitem[Karlsson et al.(2004)]{kar04} Karlsson, E., Aalto, S.,
	\& Bergman, P.  2004, ASP Conf. Ser., 320, 158
\bibitem[Larson (1974)]{lar74} Larson, R. B. 1974, \mnras, 169, 229
\bibitem[Lehnert \& Heckmand(1996)]{leh96} Lehnert, M. D.,
	\& Heckman, T. M.  1996, \apj, 462, 651
\bibitem[Marcolini et al.(2005)]{mar05} Marcolini, A., Strickland, D. K.,
	D'Ercole, A., Heckman, T. M., \& Hoopes, C. G.  2005, \mnras, 362, 626
\bibitem[Martin(1998)]{mar98} Martin, C. L.  1998, \apj, 506, 222
\bibitem[Matsushita et al.(2005)]{mat05} Matsushita, S., Kawabe, R.,
	Kohno, K., Matsumoto, H., Tsuru, T. G., \& Vila-Vilar\'o, B.  2005,
	\apj, 618, 712
\bibitem[Matsushita et al.(2000)]{mat00} Matsushita, S., Kawabe, R.,
	Matsumoto, H., Tsuru, T. G., Kohno, K., Morita, K.-I., Okumura, S. K.,
	\& Vila-Vilar\'o, B.  2000, \apj ,545, L107
\bibitem[McCray \& Kafatos(1987)]{mcc87} McCray, R. \& Kafatos, M.
	1987, \apj, 317, 190
\bibitem[Nakai et al.(1987)]{nak87} Nakai, N., Hayashi, M., Handa, T.,
	Sofue, Y., Hasegawa, T., \& Sasaki, M.  1987, \pasj, 39, 685
\bibitem[Neininger et al.(1998)]{nei98} Neininger, N., Gu\'elin, M.,
	Klein, U., Garc\'ia-Burillo, S., \& Wielebinski, R.  1998,
	\aap, 339, 737
\bibitem[Okumura et al.(2000)]{oku00} Okumura, S. K., et al.  2000,
	\pasj, 52, 393
\bibitem[Rand(2000)]{ran00} Rand, R. J.  2000, \apj, 535, 663
\bibitem[Rose(1998)]{ros98} Rose, W. K.  1998, Advanced Stellar Astrophysics
	(Cambridge: Cambridge University Press)
\bibitem[Sakamoto et al.(2006)]{sak06} Sakamoto, K., et al.  2006, \apj,
	636, 685
\bibitem[Sakamoto et al.(1995)]{sak95} Sakamoto, K., Okumura, S.,
	Minezaki, T., Kobayashi, Y., \& Wada, K.  1995, \aj, 110, 2075
\bibitem[Sanders et al.(2003)]{san03} Sanders, D. B., Mazzarella, J. M.,
	Kim, D.-C., Surace, J. A., \& Soifer, B. T.  2003, \aj, 126, 1607
\bibitem[Strickland et al.(2004)]{str04} Strickland, D. K., Heckman, T. M.,
	Colbert, E. J. M., Hoopes, C. G., \& Weaver, K. A.  2004, \apj, 606, 829
\bibitem[Strickland \& Stevens(2000)]{str00} Strickland, D. K.,
	\& Stevens, I. R.  2000, \mnras, 314, 511
\bibitem[Sunada et al.(1994)]{sun94} Sunada, K., Kawabe, R.,
	\& Inatani, J.  1994, Int. J. Infrared Millimeter Waves, 14, 1251
\bibitem[Taramopoulos et al.(2001)]{tar01} Taramopoulos, A., Payne, H.,
	\& Briggs, F. H.  2001, \aap, 365, 360
\bibitem[Tarchi et al.(2004)]{tar04} Tarchi, A., Greve, A., Peck, A. B.,
	Neininger, N., Wills, K. A., Pedlar, A., \& Klein, U.  2004, \mnras,
	351, 339
\bibitem[Tarchi et al.(2000)]{tar00} Tarchi, A., Neininger, N., Greve, A.,
	Klein, U., Garrington, S. T., Muxlow, T. W. B., Pedler, A.,
	\& Glendenning, B. E.  2000, \aap, 358, 95
\bibitem[Tenorio-Tagle \& Bodenheimer(1988)]{ten88} Tenorio-Tagle, G.,
	\& Bodenheimer, P.  1988, \araa, 26, 145
\bibitem[Tomisaka \& Ikeuchi(1988)]{tom88} Tomisaka, K.,
	\& Ikeuchi, S.  1988, \apj, 330, 695
\bibitem[Tsutsumi et al.(1997)]{tsu97} Tsutsumi, T., Morita, K.-I.,
	\& Umeyama, S.  1997, in ASP Conf. Ser. 125, Astronomical Data
	Analysis Software and Systems VI, ed. G. Hunt \& H. E. Payne
	(San Francisco: ASP), 50
\bibitem[Tully(1988)]{tul88} Tully, R. B.  1988, Nearby Galaxeis Catalog
	(Cambridge: Cambridge University Press)
\bibitem[Walter et al.(2004)]{wal04} Walter, F., Dahlem, M.,
	\& Lisenfeld, U.  2004, \apj, 606, 258
\bibitem[Weaver et al.(1977)]{wea77} Weaver, R., McCray, R., \& Castor, J.
	1977, \apj, 218, 377
\bibitem[Weiss et al.(2001)]{wei01} Weiss, A., Neininger, N.,
	H\"uttemeister, S., \& Klein U.  2001, \aap, 365, 571
\bibitem[Weiss et al.(1999)]{wei99} Weiss, A., Walter, F., Neininger, N.,
	\& Klein, U.  1999, \aap, 345, L23
\bibitem[Wills et al.(1999)]{wil99} Wills, K. A., Redman, M. P.,
	Muxlow, T. W. B., \& Pedlar, A.  1999, \mnras, 309, 395
\bibitem[Young et al.(1988)]{you88} Young, J. S., Claussen, M. J.,
	Kleinmann, S. G., Rubin, V. C., \& Scoville, N.  1988, \apjl, 331, L81
\end{thebibliography}
\end{document}